\documentclass[12pt,preprint]{aastex}
\usepackage{amsmath, amsthm, amssymb}
\def\hb{{\sc{H}}$\beta$\/}  \def\ltsima{$\; \buildrel  <  \over \sim
\;$}    \def\simlt{\lower.5ex\hbox{\ltsima}}    
\def\gtsima{$\;      \buildrel      >      \over      \sim      \;$}
\def\simgt{\lower.5ex\hbox{\gtsima}}      

       \def\ergss{ergs
s$^{-1}$\/}

\begin{document} 
  
\title{The Compact, Conical, Accretion-Disk Warm Absorber of the
  Seyfert 1 Galaxy NGC 4051 and its Implications for
  IGM-Galaxy Feedback Processes} 
\author{Yair Krongold$^{1,2}$, Fabrizio Nicastro$^{2,1}$, Martin Elvis$^{2}$, 
Nancy Brickhouse$^{2}$, Luc Binette$^{1}$, Smita Mathur$^{3}$ \& Elena Jim\'enez-Bail\'on$^{4}$
}

\altaffiltext{1}{Instituto de Astronomia, Universidad Nacional 
Autonoma de Mexico, Apartado Postal 70-264, 04510 Mexico DF, Mexico.} 
\altaffiltext{2}{Harvard-Smithsonian Center for Astrophysics, 60 Garden 
Street, Cambridge MA 02138, USA.}
\altaffiltext{3}{Ohio State University, 140 West 18th Avenue,
  Columbus, OH 43210, USA.}  
\altaffiltext{4}{Dipartimento de Fisica,  Universit\'a degli  Studi 'Roma
  Tre', Via della Vasca Navale, 84, I-00146-Roma, Italy.}


\begin{abstract}

Using a 100~ks XMM-Newton exposure of the well studied, low black hole
mass (M$_{BH} = 1.9 \times 10^6$ M$_{\odot}$) Narrow Line Seyfert 1
NGC 4051, we show that the time evolution of the ionization state of
the X-ray absorbers in response to the rapid and highly variable X-ray
continuum
constrains all the main physical and geometrical
properties of an AGN {\em Warm Absorber} wind. We use a
technique that takes advantage of the complementary high resolution 
in the RGS grating data and the high S/N in the EPIC CCD data. The
absorber consists of two different ionization components, with a difference of
$\approx$~100 in ionization parameter, and a difference of $\approx$ 5 in column
density. By tracking the response in the opacity of the gas in each
component to changes in the
ionizing continuum, we were able to constrain the electron density of the
system. We find $n_e =(5.8-21.0) \times 10^6$~cm$^{-3}$ for the high ionization absorber and $n_e > $8.1$\times$10$^7$~cm$^{-3}$
for the low ionization absorber. Combined with the X-ray luminosity
and ionization parameter, these densities require that the high and low
ionization absorbing components of NGC~4051 must be compact, at
distances 0.5-1.0~l-d
(2200 - 4400R$_s$) and $<$ 3.5~l-d ($<~15800$R$_s$) from the continuum
source, respectively. This rules out an origin
in the dusty obscuring torus, as the dust sublimation radius is 
at least an order of magnitude larger ($\sim$12~l-d), and also rules out an
association with the low ionization H$\beta$ emitting broad emission
line region (radius 5.6~l-d).  An accretion disk origin for the warm
absorber wind is strongly suggested, and an association with the high
ionization, HeII emitting, broad emission line region (radius
$<$2~l-d) is possible.
The warm absorber has a relative thickness,
$\Delta R / R \sim$ 10\% - 20\%, and the two detected phases are
consistent with pressure equilibrium, which suggests that the absorber
consists of a two phase medium. A radial flow in a spherical geometry
is unlikely, and a conical wind geometry is preferred.
The implied mass outflow rate from this wind, can be well
constrained, and is $2-5$ \% of the mass accretion rate. 
If the mass outflow rate scaling with accretion rate is representative
of all quasars, our results imply that warm absorbers in 
powerful quasars are unlikely to produce important evolutionary effects on their
larger environment, unless we are observing the winds before they
get fully accelerated. Only in such a scenario can AGN winds
be important for cosmic feedback.

\end{abstract}


\section{Introduction \label{intro}}

Feedback from quasars and their less luminous cousins Active Galactic Nuclei 
(AGN) - via radiation, highly directional relativistic jets, and slower wide 
angle winds - has been recognized as a potentially crucial input to 
cosmology  (e.g. Ciotti \& Ostriker 1997; Magorrian et al. 1998; Gebhardt et
al.; 2000; Ferrarese \& Merritt 2000; Elvis et al. 2002;
Hopkins et al. 2005; Nulsen et
al. 2005). However, while the first two mechanisms have 
been known for a long time, the prevalence and strength of quasar winds is 
only now becoming clear. Without a firm physical grasp of the wind
mechanisms, their importance must remain speculative. The most
fundamental  question is where do quasar winds originate?
Proposals span a factor $>~10^6$ in radius and include (a) the Narrow Line
Region\footnotemark (e.g. Kinkhabwala et al. 2002; Ogle et al. 2004), (b)  the inner
edge of the `obscuring
torus' (Krolik \& Kriss 2001) ,
and (c) the accretion disk itself (Elvis 2000). Discriminating among these
widely different scales ($\sim 10$ kpc to $\sim 0.001$ pc) requires
the independent determination of a quantity which is not directly
observable: the electron density $n_e$ of the outflowing material,
which then gives the distance $R$ from the central ionizing source. 

\footnotetext{We note that while extended and blueshifted narrow emission lines have
  been observed in Seyfert 2 Galaxies, it is not clear yet where
  these outflows (that extend for tens of parsecs away from the nucleus)
  originate. It is not clear either that these
  outflows are the same systems that form the ionized absorbers in
  Seyfert 1s.  In \S \ref{geomb} we will argue that the narrow
  emission lines and the absorption could arise from the same wind,
  but at very different locations.}

Ionized absorption outflows ({\em Warm Absorbers}: WA) have been observed in 
the Ultraviolet (UV) and X-ray spectra of $\simgt 50$ \% of Seyfert 1s (e.g. 
Crenshaw, Kraemer \& George 2003) and quasars (Piconcelli
et al. 2005), with line of sight velocities of the order of 
a few hundreds to $\sim 2000$ km s$^{-1}$. Such high detection rates,
combined with evidence for transverse flows (Mathur et al. 1995;
Crenshaw, Kraemer \& George 2003, Arav 2004) suggest that WAs are actually
ubiquitous in AGN, but become directly
visible in absorption only when our line of sight crosses the
outflowing  material. 
Recent efforts to understand WAs have concentrated on the accurate
spectral modeling of the 
hundreds of bound-bound and bound-free transitions directly visible in 
the time-averaged, high signal-to-noise (S/N), high-resolution X-ray
spectra of nearby Seyferts (e.g.  Krongold et al. 2003; Netzer et
al. 2003). While these studies have been decisive
in showing that there are just a few distinct physical components in these 
outflows, only average estimates of the product $(n_e R^2)$ could be derived 
from such time-averaged spectral analyses. This is due to the intrinsic 
degeneracy of $n_e$ and $R$ in the equation that defines the two observables: 
the average ionization parameter of the gas $U_x = Q_x / (4 \pi R^2 c n_e)$ 
and the luminosity of ionizing photons $Q_x$. 

It was recognized some years ago (Krolik \& Kriss 1995; Reynolds et
al. 1995; Nicastro et al. 1999) that there is
an unambiguous method to remove this degeneracy: by monitoring the
response of the ionization state of the gas in the wind to changes of
the ionizing continuum, it is possible to measure the density of the
gas and, hence, its distance from the ionizing source. Since the
ionization parameter depends linearly on the ionizing flux, it needs
to be assumed that the density and location of the absorber
do not change on the typical timescales of variability.
This method has been applied in the past to constrain the location of
the absorber. For instance Reynolds et al. (1995) applied this technique
to ASCA data of MCG -6-30-15 and concluded that the absorber should be
located at subparsec distances from the central source. Nicastro et
al. (1999) concluded, based on ROSAT data, that the absorber in NGC
4051 should be located at a
distance $\sim$3 l-d from the supermassive black hole (\S
\ref{nic1}). Netzer et al. (2002) used {\em Chandra} and ASCA data of
NGC 3516 to place the absorber at subparsec distances from the
ionizing source. Netzer et al. (2003) and Behar et al. (2003) used the
lack of variability in timescales of days for the warm absorber in NGC
3783, and set a lower limit $\sim$1-3 pc on the distance of the warm
absorber, while Krongold et al. (2005) reported variations in this
source on timescales $\sim$ 1 month, and placed the absorber within 6
pc. Reeves et al. (2004) further reported variability of the Fe XXV
and XXVI absorption lines in NGC 3783 and located the gas within 0.1
pc from the supermassive black hole.

Other methods have also been used to measure the location of the
absorbing gas. Kaastra et al. (2005) reported the detection of an O V
absorption line arising from a meta-stable level in Mrk 279, and used
it to constrain the
density of the gas. These authors placed the absorber at distances
between 1 light week to
few light months from the central source. Gabel et al. (2005) detected
meta-stable absorption from C III in one of the three absorption
components present in the UV data of NGC 3783 (the highest velocity
component) and place this UV absorber at a distance of 25 pc. He also
concluded that the gas producing the other two UV velocity components
must be located anywhere within 25 pc (the UV absorbers have been associated with
the X-ray absorber, Gabel et al. 2003).    
Such different determinations of the absorber location may be reflecting
simply the fact that we are observing a large scale
outflow, with different components. However, we note that to determine
the origination radius of the WA winds it is always the smallest
radius found the one that gives the strongest constraint.

Here, we follow the response of the absorbing gas in NGC 4051 to rapid
changes in the continuum, using high S/N XMM-{\em
Newton} data of this source, in combination with a detailed
multi-phase ionized absorber spectral model (PHASE,  Krongold et
al. 2003) and the constraints provided by time-evolving
photoionization (Nicastro et al. 1999), to produce an accurate 
determination of the absorber density and distance.
NGC 4051 is a low luminosity ($L_{bol} = 2.5 \times 10^{43}$ \ergss,
Ogle et al. 2004), low black hole mass (M$_{BH} = 1.9 \times 10^6$ M$_{\odot}$, Peterson et
al. 2004) AGN, which  varies rapidly ($\sim$ 1 hour) and with large amplitude (a factor $\sim$10; 
McHardy et al. 2004). We use a novel
technique that takes great advantage of the complementary high resolution 
in the RGS grating data and the high S/N in the EPIC CCD data, to effectively
track the changes in the WA properties over small time scales.

\section{The Variable Spectrum of NGC 4051 \label{dat}}

\begin{figure}
\plotone{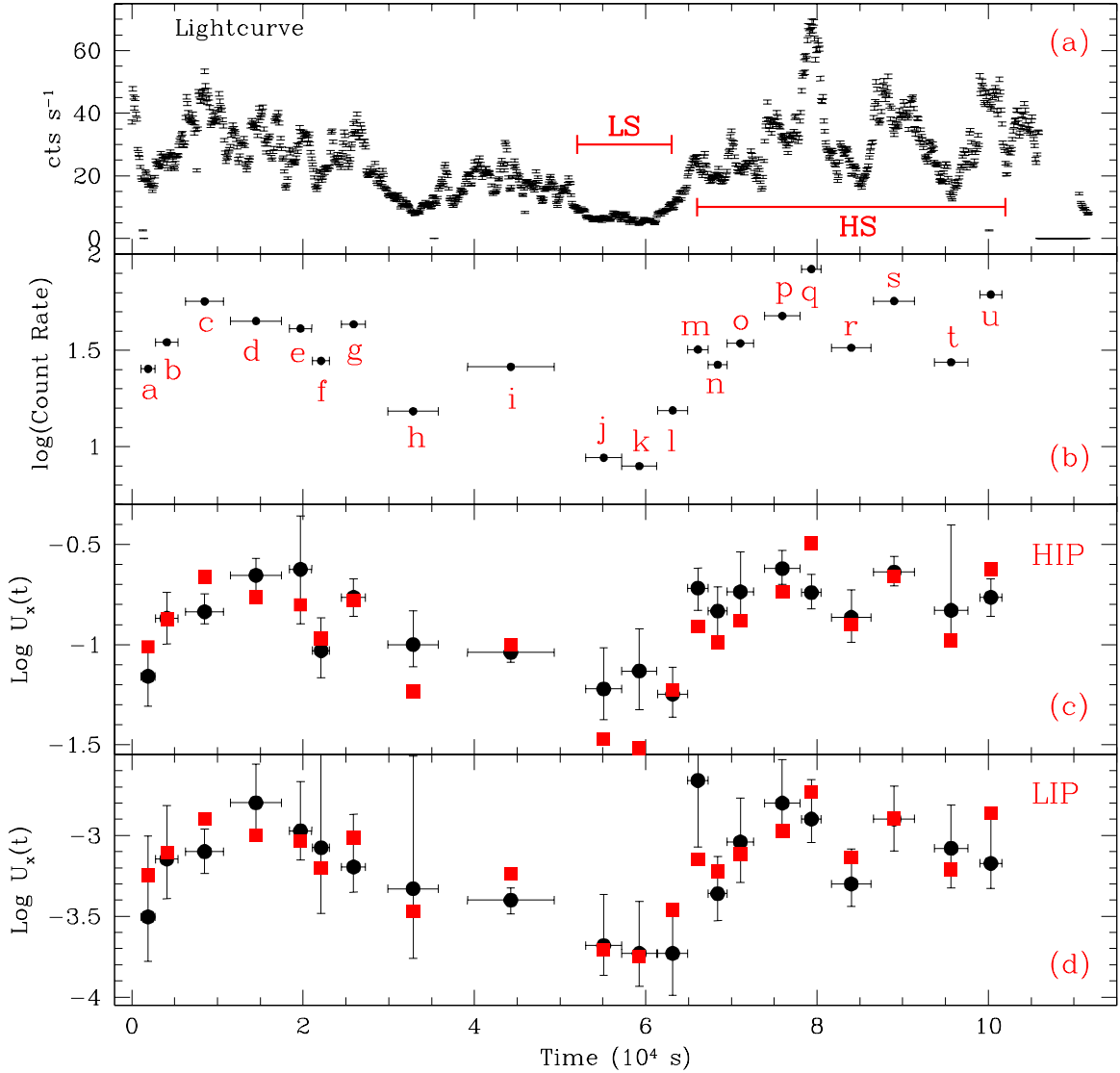} \caption[f1.eps]{
\footnotesize{
Panel (a): Lightcurve of NGC 4051 in
bins of 100 s. The high flux state (HS) and low flux state (LS)
regions used for our analysis are labeled.  Panel (b): Log of the
count rate vs. time, for the 21 ``flux states'' used to study the
variability of the warm
absorber. Spectra were extracted for each of the 21 bins, with an
exposure time given by the x error bars. Panel (c): Log of the Ionization
parameter of the HIP (U$_x^{HIP}$) as a function of time. For easy comparison with
the continum level, the red squares represent the count rate in each
bin shifted by an offset of -0.81. Panel (d): Log of the Ionization parameter of
the LIP(U$_x^{LIP}$). The red squares have an offset level of -3.05.}
}\label{lcurve}
\end{figure}

NGC 4051 was observed for $\sim$117 ks with the XMM-Newton
Observatory, on 2001 May 16-17 (Obs. Id. 0109141401). The source
varied by a factor of a few on 
timescales as short as 1~ks, and  by a factor of $\sim$12 from minimum
to maximum flux over the whole observation (see Fig. \ref{lcurve}a).
We retrieved the data from the XMM-{\em Newton} data 
archive\footnotemark 
\footnotetext{URL: http://xmm.vilspa.esa.es/external/xmm\_data\_acc/xsa/index.shtml} 
and reprocessed the observations using the XMM 
Newton Science Analysis System (SAS v6.1.0).

The task {\tt epchain} was used to obtain
calibrated event lists for the EPIC-PN camera (Struder et al. 2001) which
has a better calibration below 0.8 keV than EPIC-MOS\footnotemark.
\footnotetext{URL: http://xmm.vilspa.esa.es/docs/documents/CAL-TN-0018-2-4.pdf}
EPIC-PN operated in
small-window mode with medium filters, ensuring a negligible level of
pile-up. The last 14 ks of the observation had a background
level significantly increased due to the presence of soft photons and
these data were not included in our analysis. We reprocessed the
RGS (den Herder et al. 2001) data and extracted spectra using the task
{\tt rgsproc}. Spectral modeling was carried out with the
Sherpa (Friedman, Doe, \& Siemiginowska 2001) package of the CIAO
software (Fruscione 2002).

\section{Modeling the Spectra of NGC 4051 \label{model1}}
\subsection{Constraining the absorber properties \label{model11}}

We modeled the RGS continuum with a complex model
consisting of a power-law plus 
two thermal components\footnotemark
\footnotetext{The presence of the second thermal component (also found
  by Pounds et al. 2004 and Uttley et al. 2004) is more
  evident in the EPIC-PN data of NGC~4051, and due to its low
  temperature (kT $\sim 0.06$ keV vs. kT $\sim 0.14$ keV for the hotter
  component, see Table \ref{tab1}) and the lower
  S/N of the RGS data at long wavelengths, this component was much
  better constrained with
  the low resolution data (\S \ref{epic1}). Thus, we fixed the
  temperature of this component in our analysis of the RGS data to the
  best fit value obtained in the analysis of the EPIC data.}
attenuated by absorption due to our own Galaxy
($N_H=1.31\times 10^{20}$ cm$^{-2}$, Elvis, et al. 1989).
Strong residuals were evident in the
spectrum ($\chi^2=2607/1842$ d.o.f.). Narrow emission
lines have been reported in the spectrum of NGC~4051 from a 2002 November
XMM-Newton observation, when the source was in a very low
state (Pounds et al. 2004). The lines have constant intensity (Pounds
et al. 2004), so we
included in our models the strongest
8 emission lines with the parameters fixed at the 2002 November
values (Table \ref{tabem}), which we determined fitting the RGS data of that
observation (the analysis of this observation will be presented in the
future). Our model for the total RGS 2001 observation was
statistically better than the previous one without
lines ($\chi^2=2456/1819$ d.o.f.), confirming
the presence of these features\footnotemark.
\footnotetext{In our modeling we assume that the gas producing the
  narrow emission lines is farther from the center
  than the ionized gas producing the absorption, i.e. we assume that
  the lines are not absorbed. This is consistent with the results
  found in \S \ref{loc}, and with the idea that the narrow
  emission lines and the absorption could arise from the same wind,
  but at very different locations (\S \ref{geomb}).}
However, the spectrum still showed
large residuals 
similar to those produced by an ionized absorber. In particular,
strong residuals were present in regions consistent with strong
absorption lines by H-like and He-like ions of O, N, and C. 

We then added to our model an ionized absorbing component
using PHASE (Krongold et al. 2003), a spectral code that has proven
highly successful in modeling absorption by photoionized gas. PHASE
has three free parameters for each component: the ionization
parameter, $U_X$, in the 0.1~keV-10~keV range (Netzer 1996);
the equivalent H column density, N$_H$, and the line of sight
outflow velocity of
the absorber, $v$. A fourth parameter, the turbulent velocity of the
gas, was set to 300 km s$^{-1}$ (see Krongold et al. 2003 for
details). Throughout the paper we have assumed solar
elemental abundances for the absorbing gas (Grevesse et al. 1993).  
The inclusion of this ionized absorbing component  significantly
improved our fits ($\chi^2=2116/1816$ d.o.f.). An F-test
indicates that the absorber is required at a significance level of 99.99\%. 
However, this model still left weak residuals in the region between 12-14 \AA\ where
Fe L-shell lines are intense, suggesting the possible presence of a
second ionization component, as has been found for other Seyfert~1 galaxies
(e.g. Steenbrugge et al. 2003; Krongold et al. 2003; Blustin et
al. 2005, Costantini et al. 2006).

\begin{figure}
\plotone{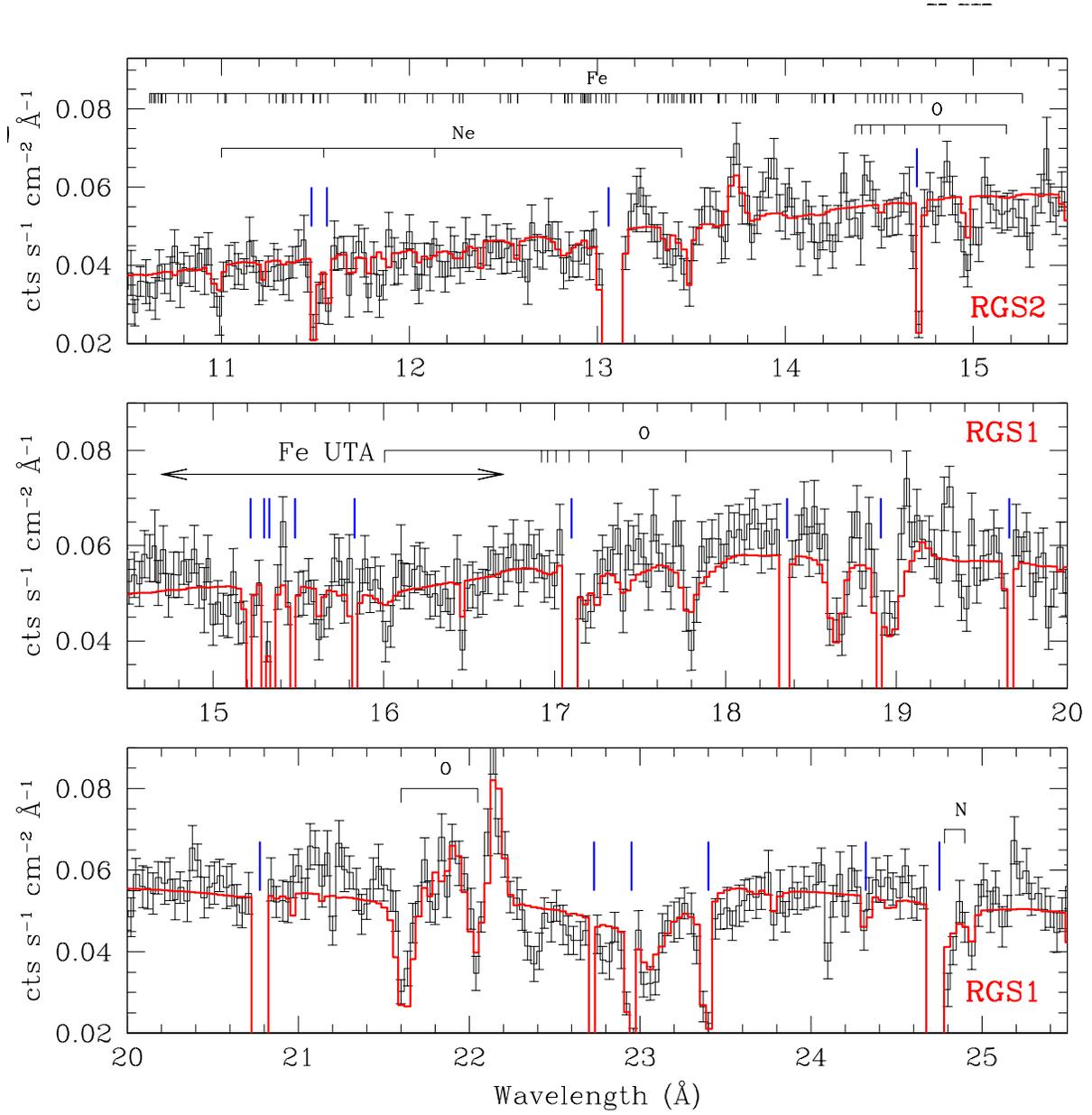}
\caption{Two absorber best-fit model over the full time integrated
RGS sepectrum. 
Absorption lines are labelled according to the element producing
them. The blue lines mark instrumental features.\label{2abs}}
\end{figure}

We thus included a second ionization component in our model and re-fit
the data. This model (reported in Table \ref{tab1})
fit better ($\chi^2=2033/1813$ d.o.f.) than the one
absorber model. An F-test
shows that the second absorbing component is required at a significance level of
99.99\%. This is consistent with results from other analyses of this source where two
absorbing components were also found (e.g. Pounds et al. 2004; Ogle et
al. 2004). One of the two absorbing components the low ionization
phase (LIP), has an ionization parameter two orders of magnitude smaller
than the other component, the high ionization phase (HIP). We will
further show in \S \ref{loc} that these two components do not
represent just a sufficient fit to the absorber (that could  instead be formed
by a continuous radial distribution of ionization parameters), but that
they are two genuinely distinct absorbing phases.
Figure \ref{2abs} shows the RGS spectrum with our best-fit model.

Finally, we note that, in modeling the same NGC 4051 spectrum, Pounds
et al. (2004) also used a description of the continuum with two thermal
components (Uttley et al. 2004 also required two thermal components
for their analysis of the EPIC data of NGC 4051). Ogle et al. (2004)
found, however, that conventional continuum emission processes cannot
explain the
``soft excess'' in the spectrum of NGC4051. However, we note that Ogle
et al. did not explore the possibility of two blackbodies in their
models. Instead, they used relativistic emission lines of the entire
O~VIII series plus radiative recombination to describe the
continuum. The nature of the soft excess observed in the X-ray spectra
of Seyfert galaxies is still not well understood. In the case of NGC 4051,
we note that the data is not sufficient to decide between two
blackbodies and relativistic emission by O~VIII. Both models provide
a good description of the continuum emission, and we point out
that using one or the other representation of the continuum has no
effect on the results found here for the ionized
absorber in NGC~4051.   

\subsubsection{On the Presence of Broad Emission Lines}

Ogle et al (2004) reported the presence of two broad emission lines (O
VII $2p-1s$ [r] and C VI  $2p-1s$ [Ly$_\alpha$]), 
presumably from the Broad Emission Line Region, in the spectrum of NGC
4051. There is indeed an excess of flux with respect to the fitted
continuum in our models, in the ranges  21-22 \AA, and 33-34 \AA\
(see also Figure 5 in Ogle et al. 2004), where these lines should be present.
We thus added two Gaussians to our model and,
according to an F-test, found that these broad features are required
with a 99.9\% level of confidence. However, despite this large statistical
significance, the lines rise only 5-10\% above the continuum
level which makes their detection unreliable. Possible residual calibration uncertainties\footnotemark,  among other effects, could in principle produce similar features.
To further test whether the emission lines are real detections or not,
we looked for possible
emission by the intercombination and forbidden O VII lines. According
to Porquet \& Dubau(2000),
in photoionized plasma, even at electron densities
$\sim10^{10}$~cm$^{-3}$ the O~VII forbidden line should be $\approx$3 times
brighter than the resonance line. Only at densities  $\sim10^{12}$
cm$^{-3}$ the forbidden line is significantly suppressed (to about
half the brightness of the resonance line). However, at these
densities, the intercombination lines become 3-4 times brighter than
the resonance line. We thus attempted to fit an O VII triplet composed
of 3 broad Gaussians, imposing these restrictions in the fluxes of the
lines. The data is not consistent with such emission line ratios,
implying no significant contribution to the emission by the forbidden
and intercombination lines (as can be seen at wavelengths larger than
22~\AA, where the low energy wing of the forbidden line should be
present). In order for the resonance line to
dominate the emission, densities much larger than $10^{12}$~cm$^{-3}$
are required. However, these lines also need a high ionization state for
the emitting gas (similar to the one of the HIP), and such large
densities would locate the gas within the horizon of events  (see \S
\ref{geom}).
Thus, it is unlikely that the flux excess between 21 and 22 \AA\ is due
to a broad O VII line.
Furthermore, 5-10\% flux excesses can be found in other regions of the
spectra where broad emission lines are not expected. We conclude that
the present data is not suitable for detecting such broad and weak
emission features.
\footnotetext{http:$\setminus\setminus$xmm.esac.esa.int$\setminus$external$\setminus$xmm\_sw\_cal$\setminus$calib$\setminus$documentation.shtml}

\subsection{Analysis of Chandra Data of NGC 4051 \label{chandra1}}

From Figure \ref{2abs}, we note that the the HIP is not as evident
(visually) in the data as the LIP 
due to several instrumental features in the RGS detector that lie
very close to the expected absorption lines, compromising their
identification. In addition, in the region between 10 and 14 \AA\, where
most of the absorption features of the HIP lie, the RGS
S/N is limited. Only the RGS 2 detector is in operation in this
region, while at higher
energies, where more HIP absorption lines are expected, the
sensitivity of the RGS declines rapidly, making the 
detection of this component difficult.

\begin{figure}
\plotone{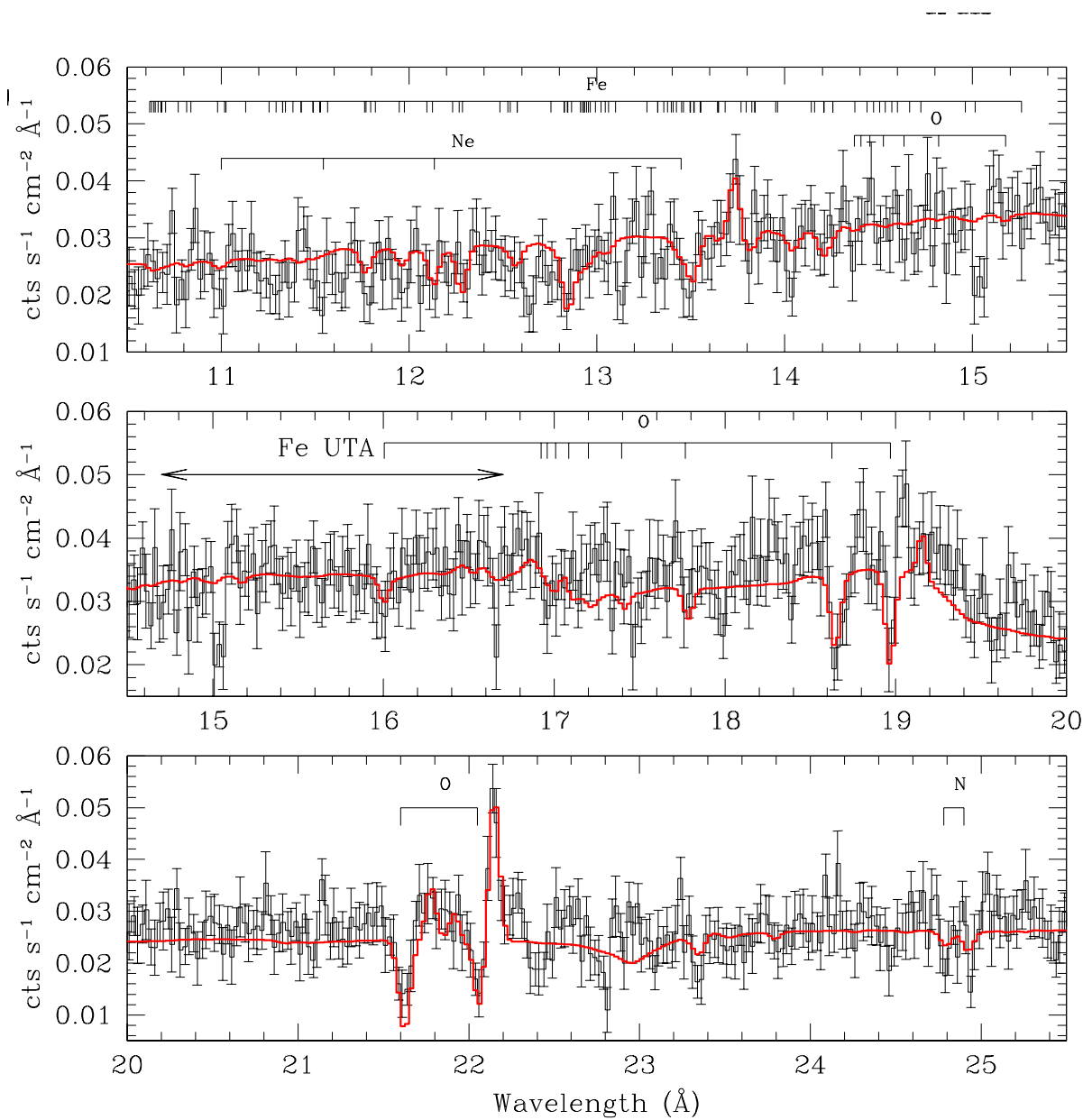}
\caption{Two absorber best-fit model over the {\em Chandra} HRC-LETG sepectrum. 
Absorption lines are labelled according to the element producing them.
The strong absorption line at 15 \AA, corresponding to an Fe XVII
transition, is clearly underpredicted by our model. \label{chandra}}
\end{figure}

As noted by Williams et al. (2006), the gratings on board {\em Chandra} have
a notably better sensitivity for detecting individual narrow absorption lines.
Thus, to further confirm the presence of the HIP we retrieved and
analyzed a $\sim 100$
ks {\em Chandra} HRC-LETG observation of NGC 4051 (a full analysis of
these data will be presented in a
forthcoming paper), carried out
in July 2003, for which the source flux is similar to that
during the XMM observation (the flux in the
6-35 \AA\ range was $\approx 3.7\times 10^{-11}$ erg cm$^{-2}$ s$^{-1}$
during the XMM observation and $\approx 4.4\times 10^{-11}$ erg
cm$^{-2}$ s$^{-1}$ during the {\em Chandra} observation).
We modeled these data following the same approach we
used with the RGS data. Our model clearly indicated the presence of
the same two components found in the RGS data, with similar column densities
and outflow velocities. The ionization parameters measured for both
components were also indistinguishable in the two spectra, implying an
absorber close to
photoionization equilibrium (log~U$_x^{LIP}$(Chandra)~=~-3.14;
log~U$_x^{HIP}$(Chandra)$~=~-0.79$). Figure \ref{chandra} presents our model
over the {\em Chandra} data. The presence of both the LIP and HIP components
is evident in the spectrum, and an F-test gives a significance larger
than 99.99\% for the presence of the second absorbing component (the HIP).

The LIP produces strong absorption
features due to O VII (lines at 21.6 \AA, 18.63 \AA, 17.77 \AA, 17.40 \AA, and
17.20 \AA) and OVI (line at 22 \AA), as well as Fe M-shell absorption
(the unresolved transition array or UTA, between 15 and 17 \AA).
The HIP  produces absorption lines 
by O VIII (at 18.97 \AA) , Ne IX-X (at 13.45 \AA, 12.13 \AA, 11.54
\AA, and 11.00 \AA), and Fe L-shell lines (between 10 and 15 \AA), in
particular by Fe XIX (intense lines at 13.80 \AA, 13.64 \AA, 13.55
\AA, 13.52 \AA, 13.50 \AA, 13.46 \AA\ and 13.42 \AA), Fe XX
(intense lines at 13.1 \AA, 12.97 \AA, 12.91 \AA,
12.85 \AA, 12.82 \AA\ and 12.57 \AA), and Fe XXI (most intense line
at 12.28\AA). 
We note that in both the RGS and  LETG data
there is a significant absorption line at 22.38$\pm$0.01 \AA\
(EW$=3.5\pm0.7$ m\AA)  that is not
reproduced by the model. This line can be identified with
an O V transition not included in PHASE. The feature at 
22.80$\pm$0.02 \AA\ (EW$=1.9\pm 1.1$ m\AA) could be an absorption line
produced by O IV also missing in PHASE, though this feature is
marginally significant in the data.


\subsection{Variability in the opacity of the ionized absorbers \label{rgsvar}}
To study the response of the absorbers to ionizing flux changes we
extracted RGS spectra during a ``low flux state'' (hereafter LS)
and a ``high flux state'' (hereafter HS). These states can be clearly
identified in Figure 1a, and they correspond to a large variation in
flux (by a factor $\sim 4.5$) over a short timescale ($\sim 5$ ks). 
Following the approach described in \S \ref{model11}, we fit
the LS and HS spectra of NGC 4051 separately, but fixing the equivalent H column
densities and outflow velocities of the absorbers to the values
obtained for the whole observation (Table \ref{tab1}). The best-fit
ionization parameters of the two absorbing
components (Table \ref{tab3}) vary between the LS and HS, 
indicating that there are spectral variations that can be fairly modeled
as opacity changes between the two states. These changes in
opacity were first noted and reported by Ogle et al. (2004) on this
same data set, however, these authors did not model the
time-dependence of the absorber as we do here.

For the LIP, the detected
change in $U_x$ between the HS and LS RGS data is significant at the
$3\sigma$ level (Table \ref{tab3}) and, within the errors, the change
in ionization parameter is similar to the change in flux (factor
$\sim$4.5). This strongly
suggests that during the flux increase from the LS to the HS, this
absorbing component reaches photoionization
equilibrium with the ionizing source.   
For the HIP the statistics are poorer. However, even for this
component a change in  $U_x$ is suggested, though significant only at the $ 1.6\sigma$
level, and is consistent with the change in flux 
within the uncertainties, allowing this absorbing component
too to be close to photoionization  equilibrium. 
For the ionized absorbers to reach photoionization equilibrium in 5
ks they must be dense and thus close to the ionizing continuum
source. We will quantify this statement in \S \ref{sec:phys}.

\begin{figure}
\plotone{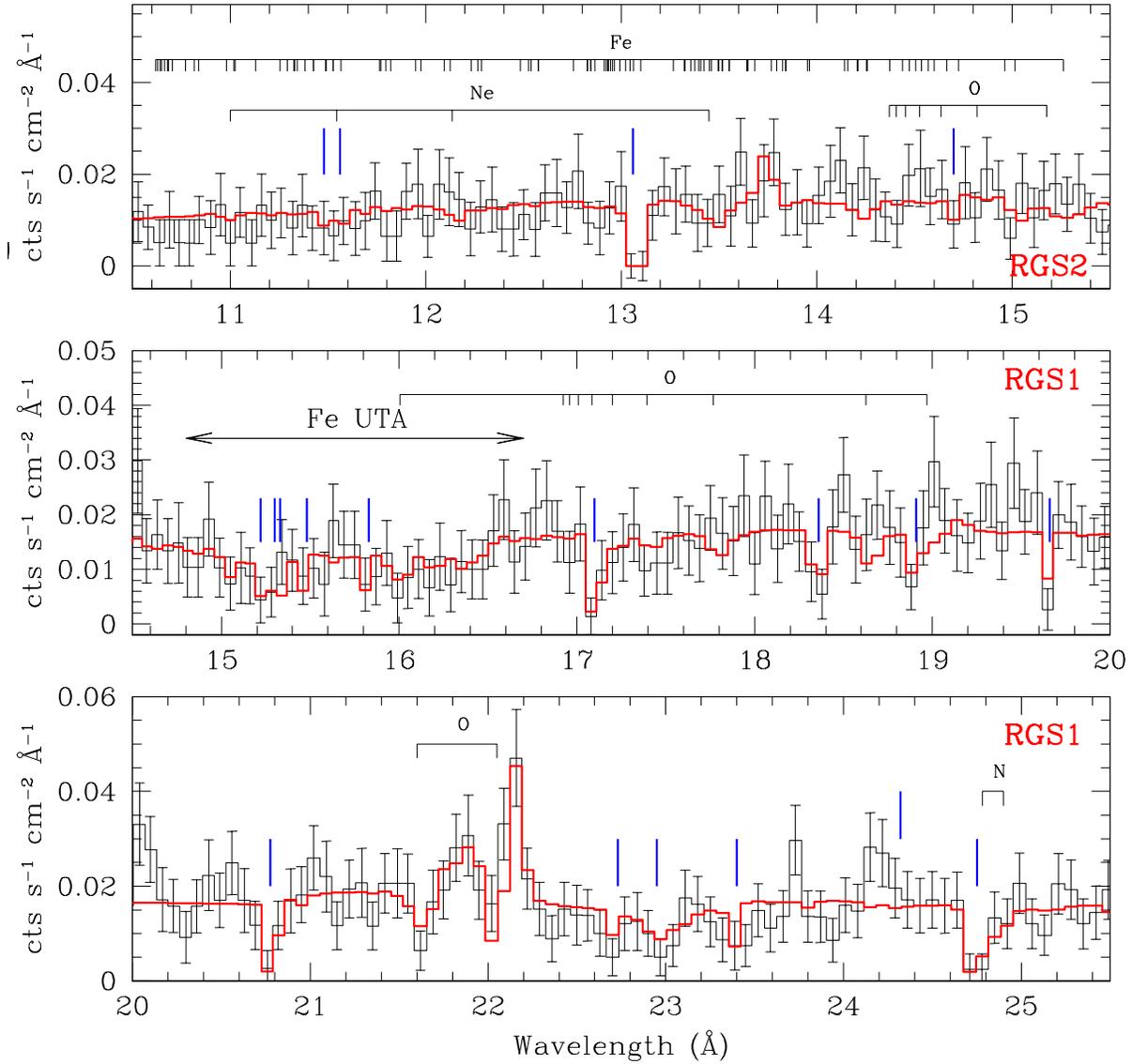}
\caption{Low State (LS) best-fit model plotted over the RGS Low State
  data. Absorption lines are labelled according to the element producing
them. The blue lines mark instrumental features.
\label{low_state}}
\end{figure}

\begin{figure}
\plotone{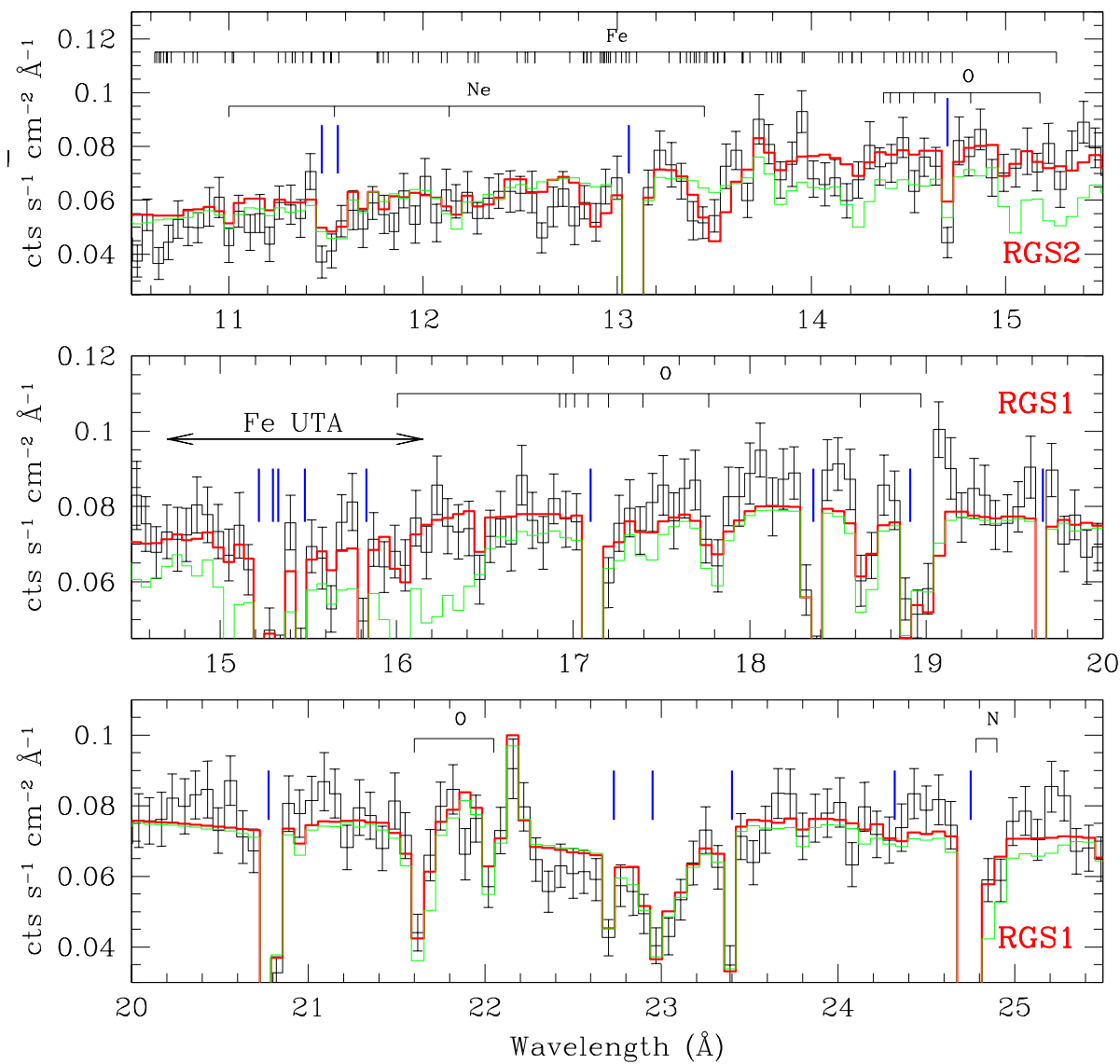}
\caption{ High State (HS) best-fit model plotted over the RGS High State
  data. Absorption lines are labelled according to the element producing
them. The blue lines mark instrumental features. The green line
  represents the best fit model obtained for the fit of the Low State RGS
  spectrum for comparison (the LS best-fit model). \label{high_state}}
\end{figure}

The best fit model for the RGS LS is presented in Figure
\ref{low_state} and the best fit model for the RGS HS is in Figure 
\ref{high_state} (in both Figs. the RGS data are binned to $\sim$55 m\AA\ per bin).
It can be observed that the variations in the spectrum of NGC~4051 are 
indeed consistent with the changes expected for the WA. In particular,
the spectral changes in the region
between 15 and 18 \AA\ can be interpreted as variations in the opacity
of the gas composing the LIP, reflected in the broad feature produced
by the Fe M-shell UTA absorption. This can be clearly seen in  Figure
\ref{high_state} with the HS data by comparing the red line (the high
state model) with the green line (the LS model)\footnotemark 
\footnotetext{ The green LS model in  Figure
\ref{high_state}
was produced using the same continuum found for the HS model, but
using the opacity produced by the absorber of the LS model. Since
the bound-free opacity of the gas is larger during the low state, the LS
model was further shifted up by 0.003 cts s$^{-1}$ cm$^{-2}$ \AA$^{-1}$ to
match the continuum level of the HS model.}. Between 10.5 and 15 \AA\ changes for
the HIP are expected, however these changes are less evident than
those for the LIP, and weaker because of the numerous blends of Fe L-shell
transitions arising from different charge states. This make the
changes for this component harder to
see by eye. However, comparing the LS and HS models in Figure
\ref{high_state}, broad changes can be observed between 14 and 15 \AA. 
Additionally, a clear variation in a semi-narrow feature
is observed in the complex around 13.5 \AA\ produced by several
transitions of Fe XIX. Another semi-narrow
feature where variation is suggested is the complex around 12.9 \AA\
produced by several transitions of Fe XX.

We note that variations in narrow absorption
features are not seen in the RGS spectra at a significant level mainly
because of (1) the limited spectral resolution of the RGS (2.4
times broader FWHM than the LETG on board {\em Chandra}, Nicastro et
al. 2007, in preparation) and (2) the low S/N ratio of the LS
data. Then, in these data only the
strongest absorption lines can be detected, but such lines are heavily
saturated, and therefore they are not expected to vary
significantly. In addition blending of weaker lines  from
different charge states, instrumental features in the RGS
coinciding with some of the most promising absorption lines, and line
absorption filling by nearby narrow emission lines complicate further
the detection of variability in narrow features. We list some of the 
non-saturated, unblended, narrow absorption lines that are expected to vary, and are
promising to be detected in high
resolution data with much higher S/N (i.e. the data expected to be
obtained with future X-ray missions like Constellation-X, Xeus or Pharos):
For the HIP Ne~X at 10.24 \AA, O VIII at 14.82 \AA, and 15.18 \AA, Fe
XVII at 10.77 \AA, 11.02 \AA, 11.13,
11.25 \AA, 13.82 \AA, 14.37 \AA, and 15.26 \AA\ are expected to be more
prominent when the source is at flux
levels similar to the one of the low state (LS). Fe XXI at 12.28 \AA,
is expected to be prominent during flux
levels similar to the HS.
For the LIP O VII at 17.20 \AA\ and 17.40 \AA, O VI at 22.03 \AA, O V
at 22.37 \AA\ (not included in our models), and N VI at 24.90 \AA\ are
expected to be more prominent during the LS.

\subsection{Modeling the EPIC data \label{epic1}} 

Motivated by the suggested 3$\sigma$ and 1.6$\sigma$ variability seen
with the RGS we analyzed the low resolution, but higher S/N, EPIC-PN spectrum of
NGC 4051. The dominance of the broad UTA and Fe-L shell features in
the RGS spectrum of the ionized absorber holds out the promise that
this low resolution  but higher S/N data could also constrain the
changes in opacity of the absorber.   

We modeled the EPIC-PN spectrum of the whole observation using the
same spectral components used to model the RGS data (including a
power-law plus two thermal components for the continuum, plus two
ionized absorbers). We fit the EPIC
spectrum leaving all parameters free
to vary, except the outflow velocity and
column density of each absorbing component, which were fixed to the
best-fit RGS values.  The best-fit parameters are listed in Table 1.

The ionization parameters derived from the EPIC data are
fully consistent with those obtained independently from the RGS data.
This result shows that when the velocities and column densities of ionized absorbers
are constrained by high
resolution data, the ionization parameter can be well
modeled at CCD X-ray spectral resolution (see also Krongold et al. 2005b).  
  
As a further check, we
extracted EPIC-PN spectra from the same high (HS) and low (LS) states
used for the RGS analysis (\S \ref{rgsvar}). The change in the
opacity of the absorbers detected in the high resolution
RGS spectra can also be detected
(though not resolved) with the higher S/N of the CCD spectra.
Quantitatively, the best-fit values of the ionization
parameters are consistent with those derived from the high resolution
data, for both absorbing components (HIP and LIP) and
during both flux states (LS and HS, see Table
\ref{tab3}). However, the values derived from the EPIC-PN data are
more tightly constrained due to the higher S/N of the CCD data (with
the velocity and column densities fixed at the RGS values). For the
LIP, the EPIC data show
opacity variations at a significance level $\sim4.3\sigma$ (Table
\ref{tab3}),  confirming the RGS
results that were significant only at a $\sim3\sigma$ level, while
for the HIP the significance level increases from the 1.6$\sigma$
level for the RGS data to a $\sim3.5\sigma$ level (Table
\ref{tab3}).
For both components the models are consistent with the gas reaching
photoionization equilibrium.

\section{Following the time-variability of the ionized absorbers}

Since NGC 4051 varies on typical timescales of a
few ks (Fig. \ref{lcurve}), the best constraints on the time dependence of the
ionized absorbers, will come from spectra in time bins of similarly short duration.
In light of the encouraging results with the EPIC-PN data, we then
performed a more finely time-resolved
spectral analysis, exploiting the better statistics of the EPIC-PN
data, with the aim of constraining the opacity variations of the
ionized absorbers on timescales as short as few ks.
We extracted EPIC-PN spectra from 21 distinct
continuum levels ({\em a - u},
see Fig. 1b), though for two intervals ({\em d} and
{\em i}) the source varied significantly on timescales
shorter than $\sim$2 ks, so these two are still only ``time
averaged spectra''. To improve the signal to noise ratio of the
21 spectra, we binned the pulse height data to have 15 original channels per bin.

\begin{figure}
\plottwo{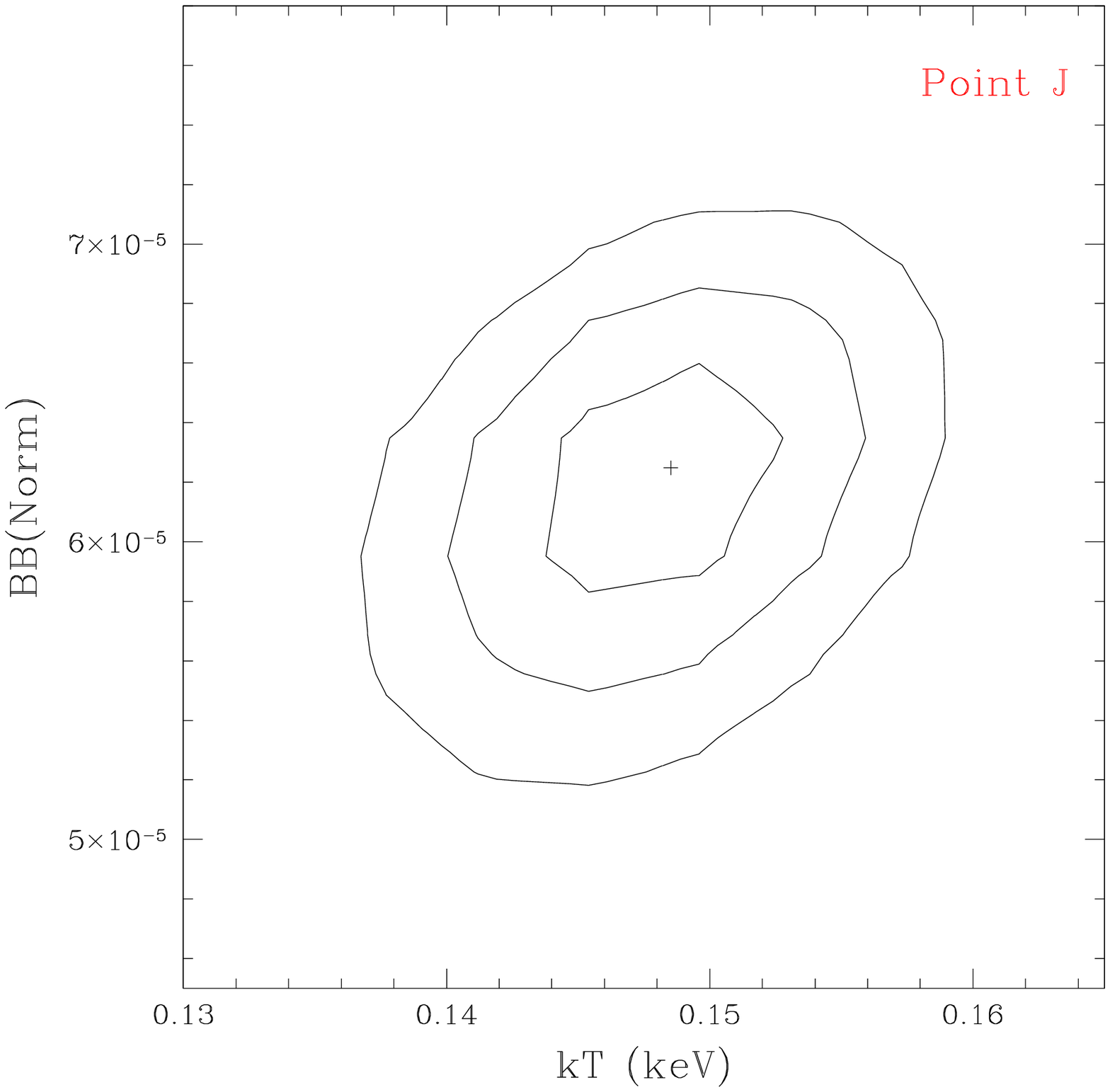}{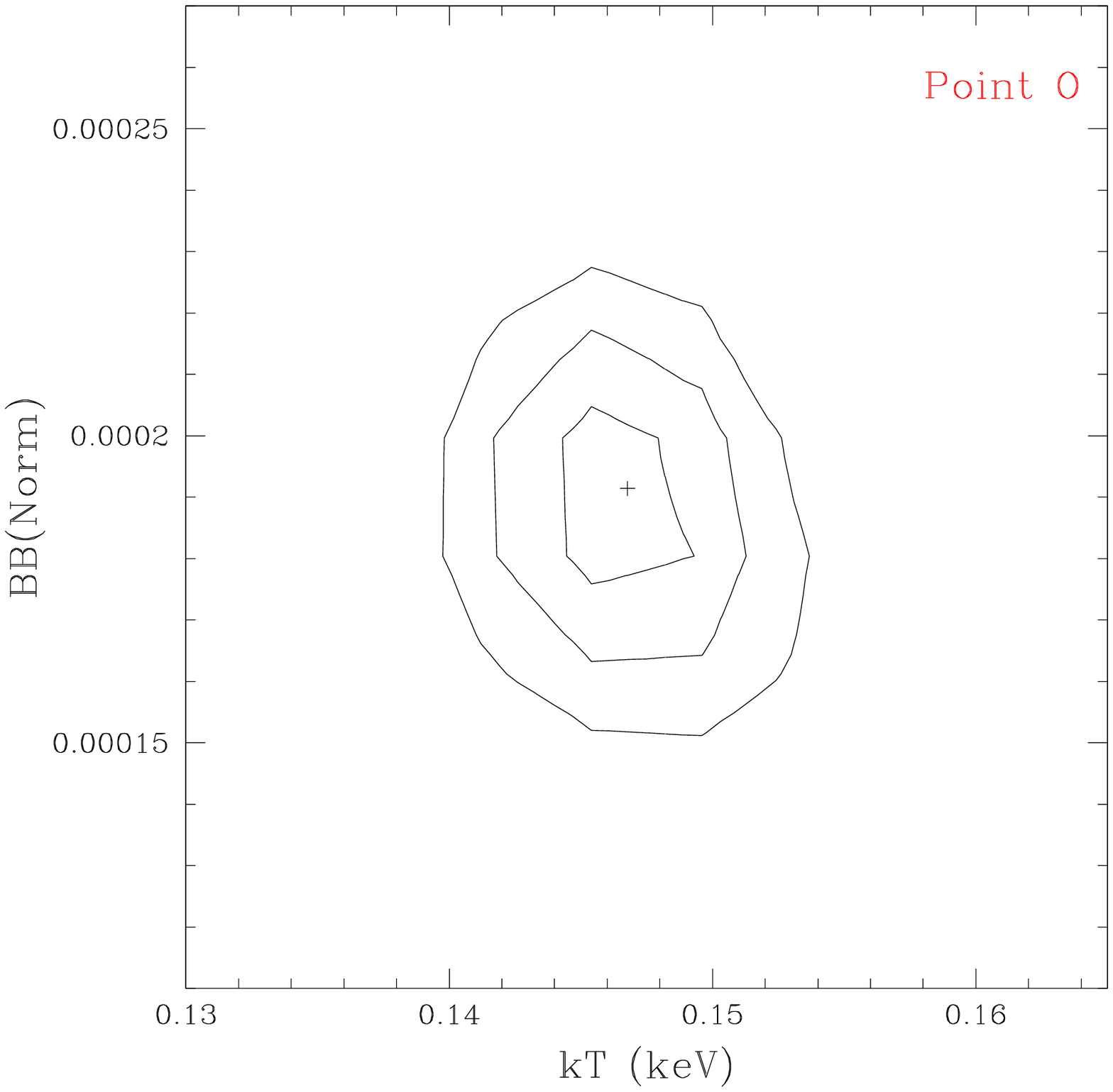}
\caption{Blackbody temperature (kT=0.14 keV) vs. blackbody normalization confidence
    regions for two spectra with
    very different flux level (states {\bf j} and {\bf o}) showing no
    correlation between these quantities. The temperature of the
    blackbody is consistent with no
    change in the 21 spectra used in our analysis. The
    normalization is in L$_{39}$/(D$_{10}$)$^2$ units, where L$_{39}$ is the
    source luminosity in units of 10$^{39}$ erg s$^{-1}$ and D$_{10}$
    is our distance to the source in units of 10 kpc.\label{tcontour}} 
\end{figure}

For each spectrum, we again fit the continuum with a power law plus
two blackbodies attenuated by both neutral absorption due to our own
Galaxy, plus the eight most prominent (non-variable) emission lines (see \S \ref{model11}). 
We also included two ionized absorbers in our models, as indicated by
the high resolution spectra.
We used the following approach: since there is
evidence that the shape of the soft excess varies only
in amplitude (see Fig. \ref{tcontour}, and Pounds et al. 2004;
Uttley et al. 2004), we required
the temperatures of the two blackbody components to be the same for
all 21 spectra. We also assumed that
the two ionized absorbers did not change in N$_H$ during the observation by
fixing N$_H$ to the same (best-fit) value in all the
models. Physically, this is plausible as a the column density of the
absorber has remained constant for several years (from the 2001
XMM-Newton observation to the 2003 {\em Chandra} observation). The
fitted N$_H$ values turned out to be consistent with the
ones determined from the RGS data. The
outflow velocity of each absorber was also assumed constant and was
constrained to the best-fit RGS value.  Thus, within the 21
individual spectra we left free to vary only: (1) the slope and
amplitude of the
power-law, (2) the amplitude of the two blackbody components,
and (3) $U_X$ for each of the two absorbers. With only 3 free
parameters, our fits are able to detect smaller changes in $U_X$.   

For both absorbing components the derived $U_X$ values  follow closely
the source continuum lightcurve [compare panel (c) and (d) with panel (b) of Figure 1], 
clearly indicating that the gas is responding quickly to the changes in the 
ionizing continuum.

\subsection{Testing the Photoionization Equilibrium Hypothesis \label{peh}}
To study more quantitatively how these changes are related to the changes in the continuum, 
we show in Fig.\ref{corr}a,b the log of 
the source count rate (log$C(t)$) vs. log$U_X(t)$, for the HIP
[panel (a)] and the LIP [panel (b)]. 
For most of the points of the HIP and for all the points of the LIP
(within 2$\sigma$), 
log$C(t)$ correlates with log$U_X(t)$ tightly. 
We then assume photoionization equilibrium and derive the quantity $(n_e R^2)$ 
by fitting the equilibrium relationship $logU_X(t) = logQ_X(t) - log(4\pi 
c n_e R^2) = log[(C(t)/S_{eff})\times (4\pi D^2)] - log(4\pi c n_e R^2)$, to 
the data points in Figure \ref{corr}. In the above formula, $Q_X$ is
the rate of 0.1-10 keV 
photons produced by the source, $(n_e R^2)$ is the product between the 
electron density of the gas and the square of its distance from the ionizing 
source, $C(t)$ and $S_{eff}$ are the observed count rate and the instrument 
effective area in the 0.1-10 keV band, respectively, and $D=15.5$ Mpc 
(Shapley et al. 2001) is the distance to NGC~4051. We adopted a value
of $S_{eff}\sim$260 cm$^2$, and performed a single parameter fit
(i.e. we assumed  $U_X = 0$ for $Q_X = 0$). Thus,  
the fits given by the solid red lines determine the
values of $n_eR^2$. We find 3.8$\pm0.7\times 10^{37} $ cm$^{-1}$ for
the HIP and
6.6$\pm0.1\times 10^{39}$ cm$^{-1}$ for the LIP (Table \ref{tab2}). 
This is a robust determination of these quantities, as it is based on 21 
different estimates of $U_X^i$ and $Q_X^i$ ($i=1, 21$), and is
consistent with the assumption of photoionization equilibrium. 
This determination of $(n_e R^2)$ is thus free from the large 
uncertainties introduced by the usual assumption of a spectral energy 
distribution for the ionizing radiation (see the discussion in
Krongold et al. 2005b).

\subsection{The Lack of Influence of the Blackbody Components on the
  Apparent Opacity Variations of the Warm Absorber \label{hsls}}

To explore whether the correlations found in Figure \ref{corr}
could be due to a bias introduced by the continuum
modeling,  rather than to real variations in the absorber,  we have run
additional tests on the 21 EPIC states {\it a-u}.

The temperatures of the blackbody
components used to model the soft X-ray continuum emission
are well constrained in all the methods we have applied, and
are independent of the way the warm absorber is modeled as shown in
Fig. \ref{tcontour} (see also Pounds et al. 2004;
Uttley et al. 2004) . Thus the temperature
parameters have a negligible influence on the values obtained
for the ionization parameters of the two absorbing
components. On the other hand, the amplitude of the hotter blackbody
component with
temperature kT$\sim$0.14 keV, might have an impact on the analysis of the warm
absorber.\footnotemark\ Since this component becomes more or less prominent
with flux
increments or decrements, the region of the spectrum where it
dominates over the power-law component also changes. Such
changes might be misinterpreted as variations in the opacities of the
absorbers. 
\footnotetext{We note that
the normalization of the blackbody component with kT$\sim$0.07 keV cannot
have an important effect on U$_X$ due to its low temperature.}

We can rule out this possibility using confidence regions for logU$_x$
vs. the normalization of the hotter blackbody component, for both the
HIP and the LIP. Figure \ref{cont_ampl} shows that, for a
representative flux state (point {\it o}), there is no
correlation between these two parameters. Variations of the
normalization of the thermal component do not influence the measured
value of logU$_x$ and, therefore, do not have an impact on the
observed correlation between flux and ionization parameter.

To further study the variations of the absorber, we have produced
confidence regions of the ionization parameter during high and low
flux states. Figure \ref{ulowvsuhigh} shows these regions for spectra
{\it j} and {\it o}. For the HIP, the variations are significant at a level
$2\sigma$. The LIP variations are significant at a level
$>3\sigma$. These tests show that changes only in the continuum cannot
account for the spectral variations, further confirming that the
absorbing material is responding to flux variations. 

We stress again that the variations we derive for U$_X$ are not random, but rather follow the
ionizing continuum level, as expected for gas close to photoionization
equilibrium. Any alternative explanation must also predict these correlations.

\subsection{Photoionization Equilibrium Timescales \label{pet}}

\begin{figure}
\plotone{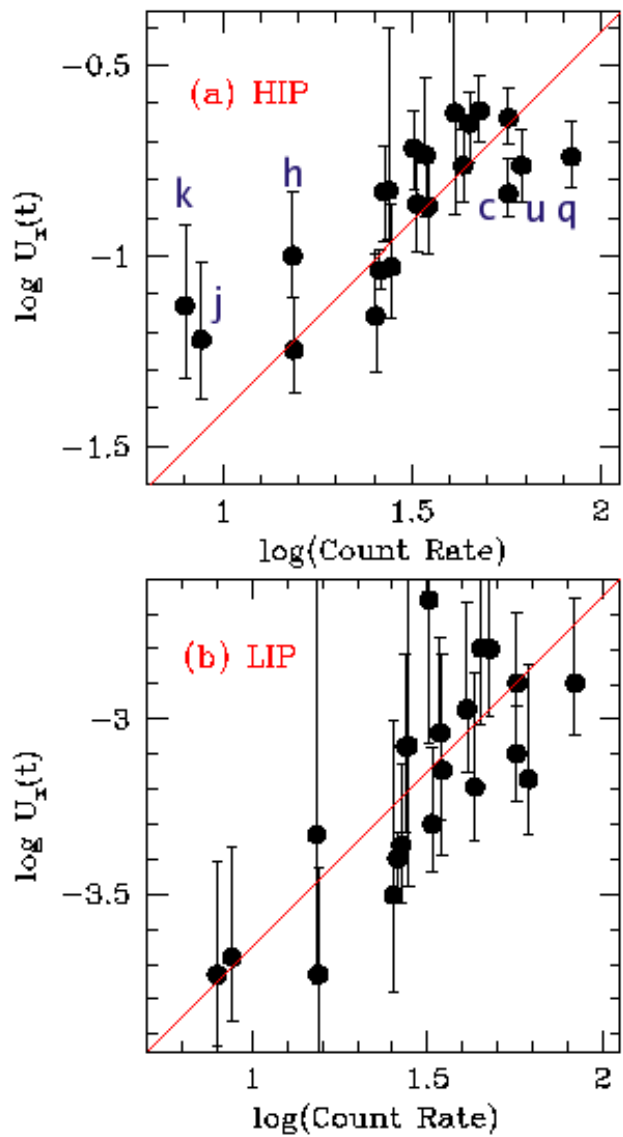} \caption[f7.eps]{Log$U_X(t)$ vs. log of the source count
rate (log$C(t)$),
for the HIP (panel (a)) and the LIP (panel (b)). 
For most of the points of the HIP and for all points of the LIP, 
log$C(t)$ correlates with log$U_X(t)$ tightly. HIP and LIP are close to
photoionization equilibrium. The solid red lines represents the
photoionization equilibrium relation and determine the values of $n_e R^2$ for 
the two components to be 3.8$\pm0.7\times 10^{37} $ cm$^{-1}$ for
the HIP and 6.6$\pm0.1\times 10^{39}$ cm$^{-1}$ for the LIP (see text for details). \label{corr}}
\end{figure}

\begin{figure}
\plottwo{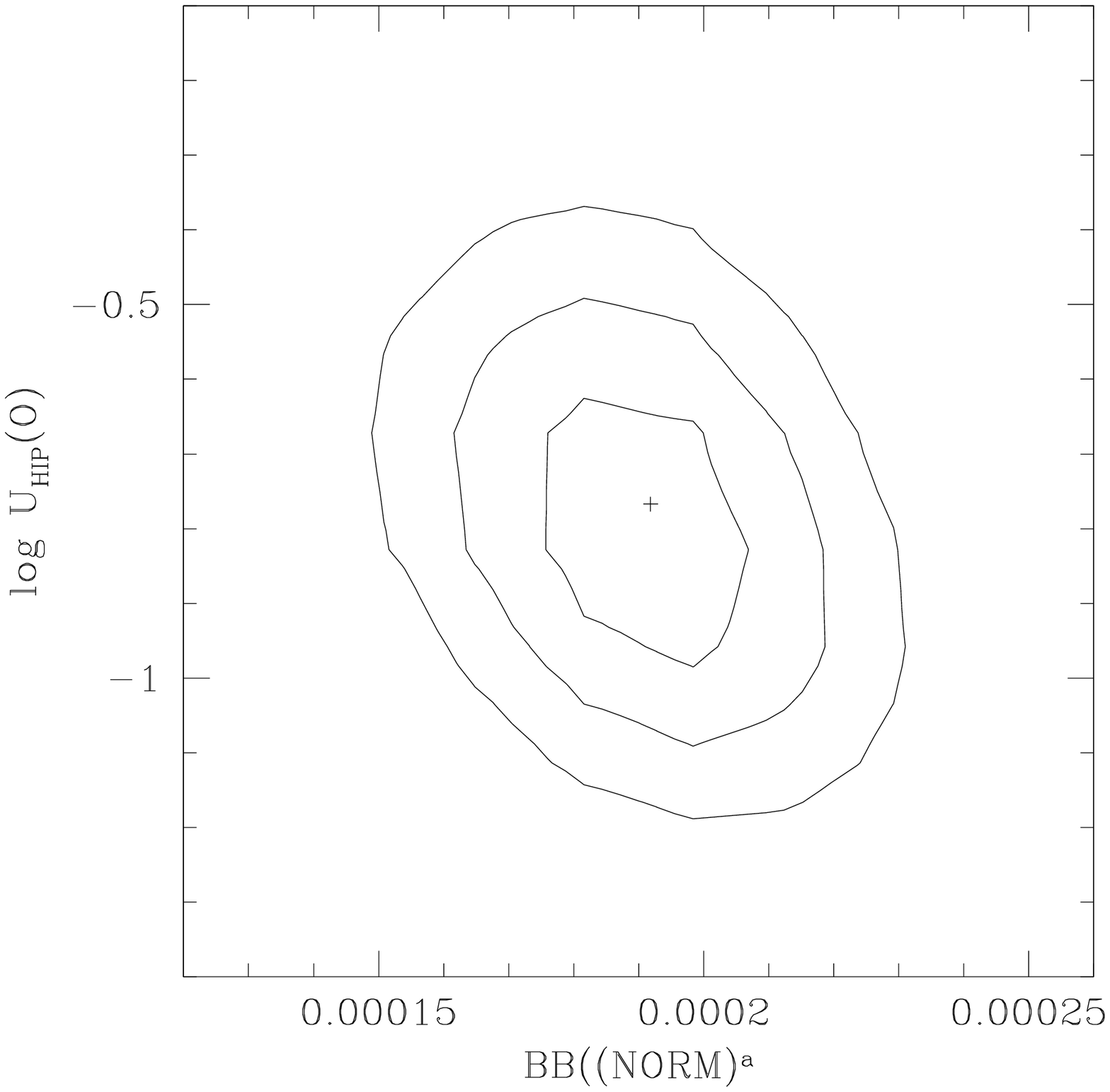}{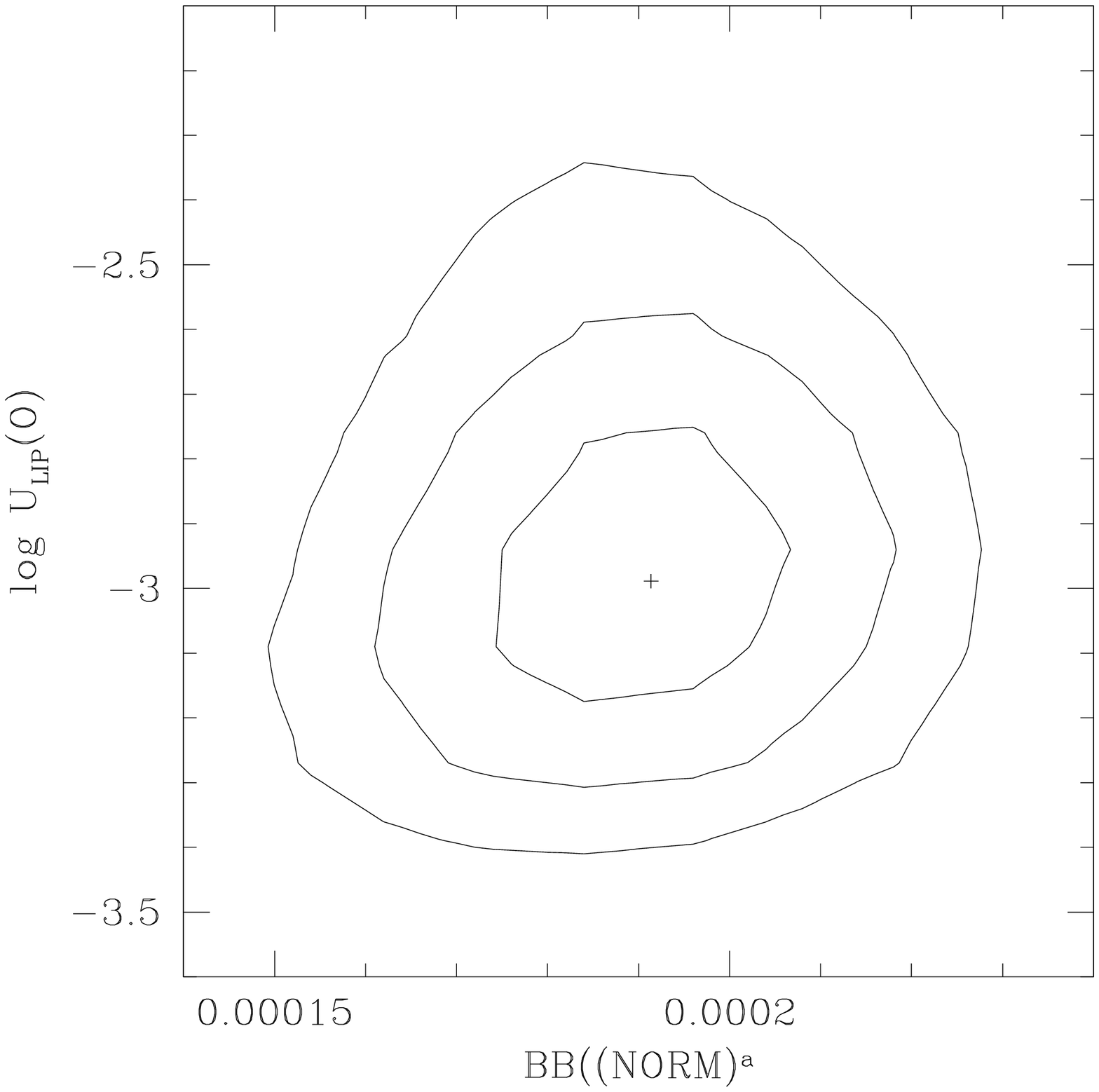}
\caption{Confidence regions for the thermal component
    normalization  (kT=.14
    keV) vs. the ionization parameter for the HIP
    and the LIP. The contours
    represent the 1, 2 and 3 $\sigma$ confidence levels. The spectra
    corresponds to time region {\it o}. There
    is no correlation between these parameters, showing that the
    continuum parameters for the blackbody do not influence the
    measured correlation U$_x$ vs. flux. Similar results are found for
    the other 20 time regions. The
    normalization is in L$_{39}$/(D$_{10}$)$^2$ units, where L$_{39}$ is the
    source luminosity in units of 10$^{39}$ erg s$^{-1}$ and D$_{10}$
    is our distance to the source in units of 10 kpc.\label{cont_ampl}}
\end{figure}

\begin{figure}
\plottwo{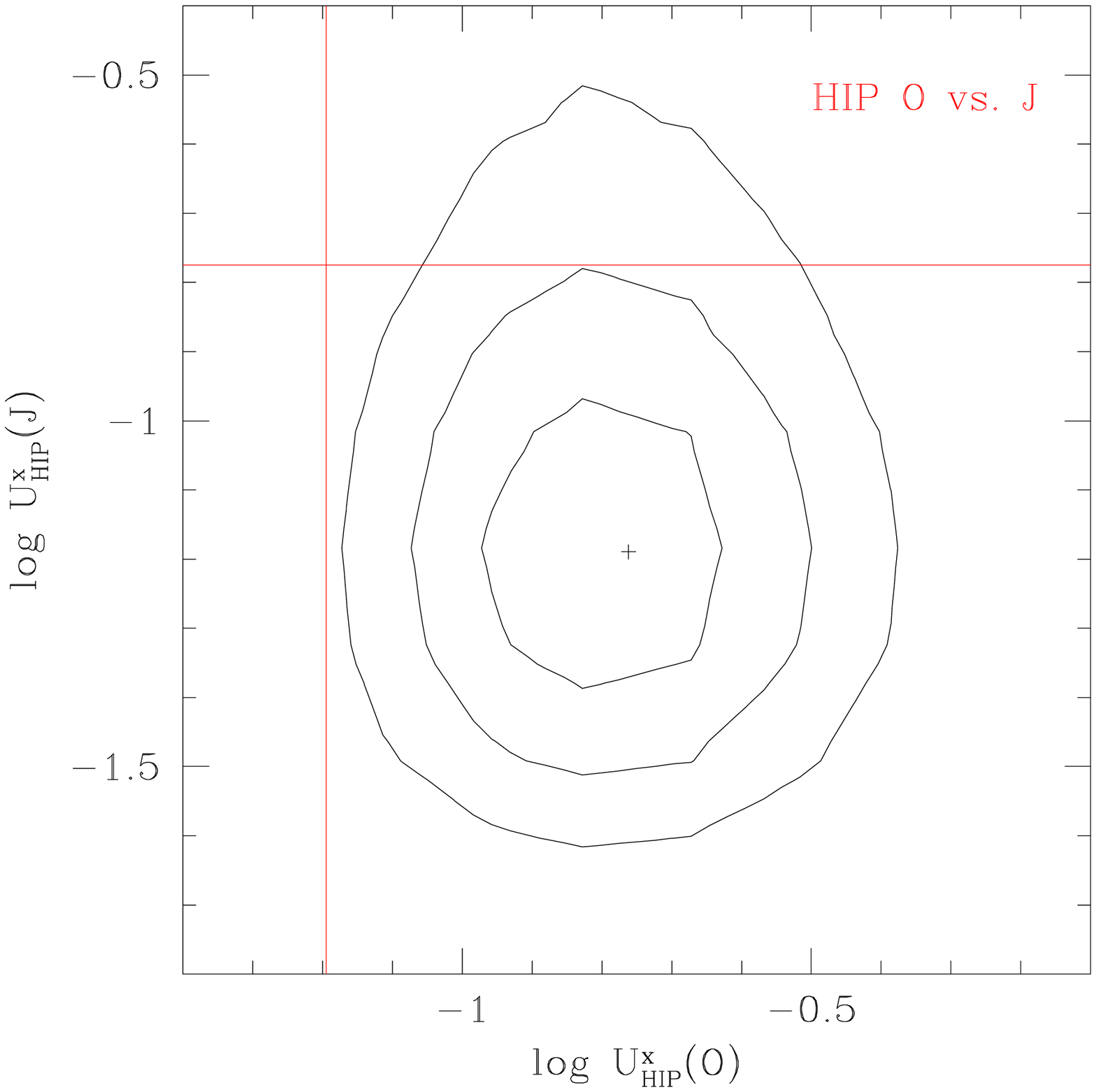}{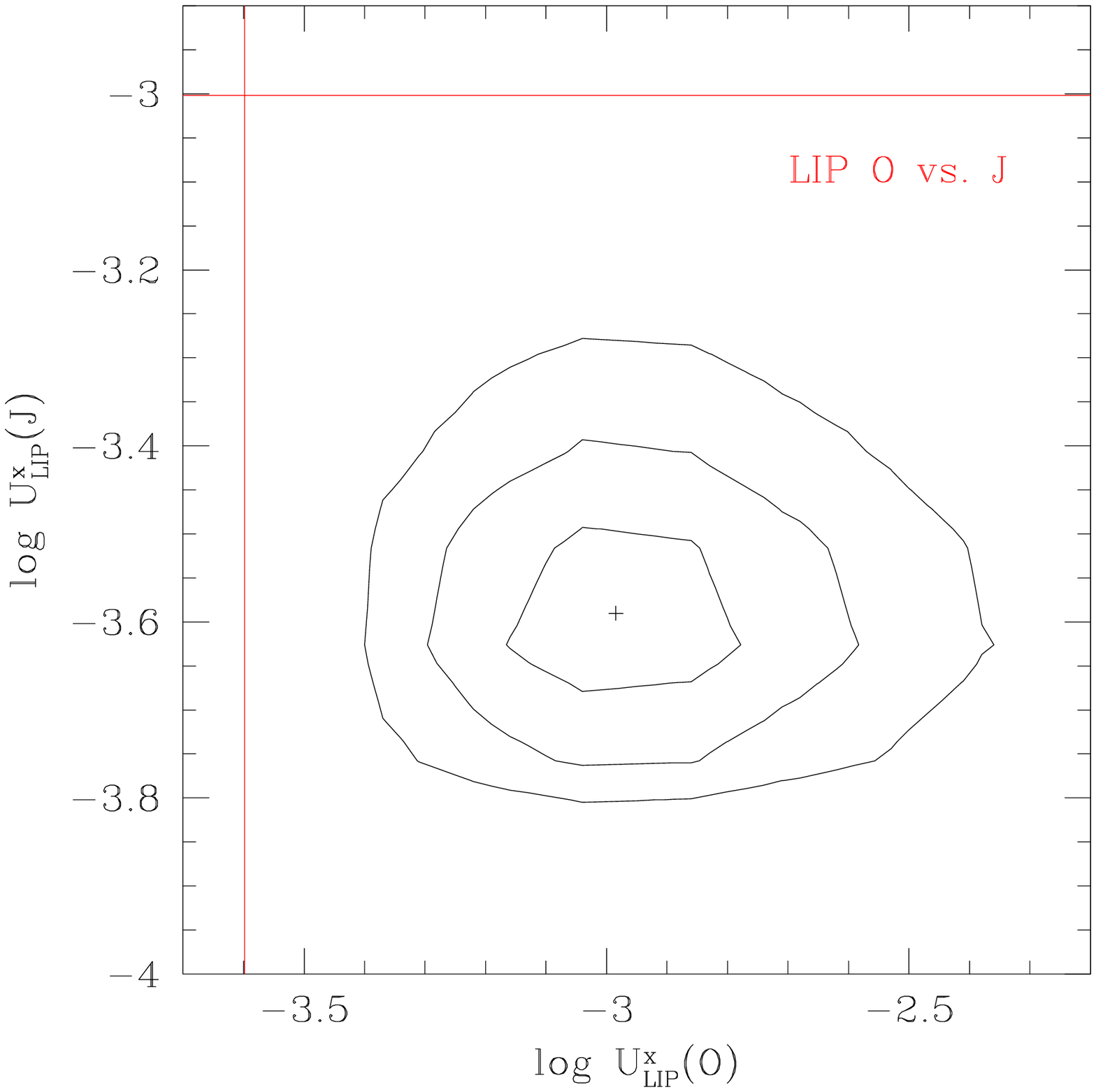}
\caption{Confidence regions for the ionization parameter during
    low and high flux levels for spectra extracted from two
    representative flux states: {\it j} and {\it
    o} of the EPIC data. We calculated the contour plots for the LIP,
    U$_x^{LIP}$({\it o}) vs. U$_x^{LIP}$({\it j}), and for the HIP,
    U$_x^{HIP}$({\it o}) vs. U$_x^{HIP}$({\it j}), thus comparing the
    state of the gas between a high and a low flux level. The contours
    represent the 1, 2 and 3 $\sigma$ confidence levels. The solid
    lines represent the ``no  variation'' lines for
    the ionization parameter [i.e. U$_x$({\it j}) $=$ U$_x$({\it
    o})]. The plots show that the
    ionization parameter is not consistent with a constant value,
    implying opacity variations. \label{ulowvsuhigh}}
\end{figure}

Since all the points for the LIP are consistent with photoionization
equilibrium within 2$\sigma$ (Fig. \ref{corr}b), the 
LIP can both recombine and ionize in a timescale shorter than the shortest 
time interval separating spectra with large flux changes. The largest
change in flux in the shortest time is between spectra
{\em l} and {\em m} (a factor of $2.1$ in flux, separated by 3~ks). 
Thus, we conclude that the {\em photoionization equilibrium timescale} (i.e. 
the time necessary for the gas to reach photoionization equilibrium with the 
ionizing source) $t_{eq}(LIP)<$3~ks.

For the HIP, the situation is somewhat more complicated. 
The spectra of the HIP at typical count-rates are all
consistent with photoionization equilibrium. 
The extreme flux points, on the other hand, show deviations from
equilibrium, i.e. they fail to respond as expected to the
changes in the continuum. 
The three lowest count rate points ({\em h}, {\em j}, and {\em k}) each lie
2$\sigma$ above the equilibrium line for the HIP in Figure \ref{corr}a
(see also Fig. 1c),
and so represent overionized
gas, i.e. gas having $U_X$ larger than expected in photoionization
equilibrium. Conversely, three of the four highest count rate points 
in Figure \ref{corr}a (see also Fig. 1c, spectra {\em c}, {\em q} and
{\em u}) lie $\sim$1-2$\sigma$ below the photoionization equilibrium
line, and so 
represent underionized gas. 
This behavior is expected in gas close to, but not instantaneously 
in, equilibrium with the ionizing flux, since, depending on gas
density, a cloud can respond quickly (few ks) to smooth and moderate
increases/decreases of the ionizing continuum, but would require
longer times to respond to fast and extreme flux
changes (Nicastro et al. 1999). These delays 
are expected to be longer during recombination (flux decreases) than during 
ionization (flux increases), implying different photoionization 
equilibrium timescales $t_{eq}$ during different source lightcurve phases. 
Unlike the LIP (for which only an upper limit on $t_{eq}$ can be 
estimated), the behavior of the HIP allows us to set both lower and upper
limits on $t_{eq}(HIP)$.

Spectra {\em j} and {\em k} correspond to a prolonged quiescent 
phase of the source lightcurve, following a smooth and $\sim 10$ ks long 
flux decrease. $U_X^j$ and $U_X^k$ are consistent with each 
other and each deviates from the equilibrium line by about 2$\sigma$
(Fig. \ref{corr}a, the two lowest count rate points). Thus, these two
spectra provide, 
when combined together, a robust lower limit on $t_{eq}(HIP)$. The total 
exposure time for spectra {\em j}, and {\em k} is $\Delta t^{j+k} \sim 10$~ks. 
In the following we then adopt $t_{eq}^{i,j+k} \simgt 10$ ks.\footnote{Superscripts on `t' refer to the time intervals labeled
in Figure 1b.} 
To estimate an upper limit on $t_{eq}(HIP)$ we use spectra {\em l} and 
{\em m} which are separated by $\sim 3$ ks. $U_X^l$ and $U_X^m$
nearly match the 
increase in flux. We conclude that, for such flux variations, the 
photoionization equilibrium time of the HIP is $t_{eq}^{l,m}(HIP)<$3~ks.\footnote{Again, we stress that $t_{eq}^{i,j+k}(HIP) \simgt 10$ ks and 
$t_{eq}^{l,m}(HIP)<$3~ks are consistent with each other, since 
in gas out of photoionization equilibrium $t_{eq}$ depends upon 
the particular phase of the ionizing source lightcurve at which such 
timescales are estimated.}

\section{The Ionized Absorber in NGC 4051: A Dense,
  Multi-Phase, Compact  Wind \label{sec:phys}}

\subsection{Physical conditions}
In \S \ref{pet}, we estimated $t_{eq}(LIP)<$3~ks (independently of the particular 
lightcurve phase, since the gas is always consistent with equilibrium) and 
$t_{eq}^{i,j+k}(HIP) \simgt 10$ ks (during a recombination phase where
the gas does not reach equilibrium) and
$t_{eq}^{l,m}(HIP)<$3~ks (during an increase in flux where the gas reaches
equilibrium with the ionizing source). 
These response times of the gas to changes in the continuum set
strong constraints on the density, $n_e$, and location, $R$, of the
absorber relative to the continuum source.

The photoionization equilibrium timescale depends on the electron 
density of the absorber, during both ionization and recombination phases 
(Nicastro et al., 1999). Thus, we can use the above estimate of $t_{eq}$
to estimate $n_e$ for the LIP and the HIP.
To obtain the densities, we used the approximate relation between
$t_{eq}$ and $n_e$ derived by Nicastro et al. (1999; eq. 5) for a
3-ion atom (i.e. an atom 
distributed mainly among three of its contiguous ion species):
\begin{equation*}
t_{eq}^{x{^i},x^{i+1}} \sim
\left[ \frac{1}{\alpha_{rec}(x^i, T_{e})_{eq}~n_e} \right] \times\left[
  \frac{1}{[\alpha_{rec}(x^{i-1}, T_e)/\alpha_{rec}(x^i, T_{e})]_{eq}
    + [n_{x^{i+1}}/n_{x^{i}}]}\right]
\end{equation*}
where ``eq'' indicates the equilibrium quantities, and
$\alpha_{rec}(x^i, T_{e}$) is the radiative recombination
  coefficient of the ion $x^{i}$, for gas with an electron
    temperature $T_e$. This is an 
excellent approximation for O and Ne for both the LIP (OVI-OVIII, NeVIII-NeX) 
and the HIP (OVII-OIX, NeIX-NeXI) because 98\% of the population of
these elements is concentrated in these charge states. We used recombination
times from Shull \& van Steenberg (1982) and the average equilibrium
photoionization temperature,
T$_{LIP}\sim 3\times 10^4$ K and T$_{HIP}\sim 5\times 10^5$ K.
For the LIP, we find $n_e(LIP)>$8.1$\times$10$^7$~cm$^{-3}$, 
and for the HIP $n_e(HIP)=(5.8-21.0) \times 10^6$~cm$^{-3}$ (Table \ref{tab2}).

The $n_e$ values allow us to obtain the gas pressure $P$ for each
component, using the average temperature.
The gas pressures of the LIP and HIP are $P_{LIP}=n_e(LIP) T_{LIP}>$2.4$
\times$10$^{12}$ K cm$^{-3}$, and $P_{HIP}$=(2.9-10.5)$ \times$10$^{12}$~K 
cm$^{-3}$, respectively (Table \ref{tab2}). The pressures of the two phases 
are therefore consistent with each other, suggesting that LIP and HIP may be 
in pressure balance, and thus two distinct phases of the same medium.
This has been suggested for several other AGNs (e.g. NGC 3783,
Krongold et al. 2003, Netzer et al. 2003; NGC 985,  Krongold et
al. 2005), but with much less tight (or no) determination of the
location of the phases (thus assuming that the two phases share the
same location).

\subsection{Location and Structure \label{geom}}

Given their densities, the distance $R$ of the LIP and the HIP from the central ionizing source can now be 
derived. We find R$_{LIP}<$8.9 $\times$10$^{15}$cm 
($<$0.0029~pc, $<$ 3.5~light-days [l-d]) and R$_{HIP}=(1.3-2.6)\times$10$^{15}$~cm 
(0.0004-0.0008~pc, 0.5-1.0~light-days [l-d]; Table \ref{tab2}). 
Thus, not only the two gas pressures, but also the distances 
of LIP and HIP from the central ionizing source, are consistent with each 
other. This gives further support to the hypothesis that LIP and HIP
are two phases of the same
wind.  We note that the opacity variations of NGC
4051, also reported by Ogle et al. (2004), rule out both a location of
parsecs from the central source and a continuous flow in our line of
sight, as suggested by these authors.

Another  lower limit on $R_{HIP}$ which is less tight, but independent
of the assumption that spectra {\em i,j+k} are out of photoionization
equilibrium, can be obtained from the ratio
$U_x(HIP)/U_x(LIP)=n_e(LIP)R^2_{LIP}/n_e(HIP)R^2_{HIP} \approx 173$.
Since the LIP responds faster than the 
HIP to changes in the ionizing flux, then $n_e(LIP)>n_e(HIP)$. 
Combining these equations gives $R_{HIP} > 0.076 R_{LIP}$ and so $R_{HIP} > 
0.27$ l-d.

The observed variability of the gas also sets constraints on the
structure of the ionized absorber. 
Assuming homogeneity in the flow, and using the column densities and
number densities inferred in our analysis, the line of sight thickness and relative 
thickness of the wind can be estimated. The line of sight thickness is given by $\Delta R \sim N_H /n_H \simeq 1.23 
N_H / n_e$ (where in the last term of the equation we used $n_e \simeq 1.23 
n_H$ which is valid for a fully ionized gas with solar abundances). 
Using simple algebra we then get the relative thickness $\Delta R / R = 1.23
N_H / n_e R = 1.23 N_H  (n_e R^2)^{-1/2} (n_e)^{-1/2}$. For our two components, we find 
$\Delta R_{LIP} < 9 \times 10^{12}$~cm, $\Delta R_{HIP}= (1.9-7.2) 
\times10^{14}$~cm, and $(\Delta R / R)_{LIP} < 10^{-3}$ and $(\Delta R / 
R)_{HIP}=(0.1-0.2)$ (Table \ref{tab2}).

Both the LIP and the HIP are then thin shells of gas (and hence they
can be analyzed accurately  in a plane-parallel configuration with
respect to the central ionizing source). Moreover,  
$(\Delta R / R)_{LIP} \simlt 1 \%$ $(\Delta R / R)_{HIP}$, suggesting that 
either the LIP is embedded in the HIP, or the LIP represents a
boundary layer of the HIP. This result is consistent with our finding 
that the thermal pressures of the two components are consistent with each 
other.

\subsection{Consistency Check with High Resolution Data}

We stress that qualitatively similar, but quantitatively less tight
conclusions on the physical state and structure
of the two absorbers can be reached based on the analysis of
the LS and HS high resolution RGS data only. The implied response time
of the warm absorber to changes in the ionizing continuum are $t_{eq}(LIP)<$15~ks 
and $t_{eq}(HIP)<$15~ks (the duration of the LS observation). This
implies $n_e(LIP)>$1.6$\times$10$^7$~cm$^{-3}$, and
$n_e(HIP)>$1.2$\times$10$^6$~cm$^{-3}$;  R$_{LIP}<$2.0 $\times$10$^{16}$cm 
and R$_{HIP}=(1.5-5.8)\times$10$^{15}$~cm ($\sim 10^{-3}$ pc). The lower limit
in R$_{HIP}$ comes from the ratio $U_x(HIP)/U_x(LIP)$. This confirms
that the warm absorber in NGC 4051 is both dense and compact.

\subsection{Comparison with Previous Observations \label{nic1}}

Using a simple time-evolving photoionization code, which included only
bound-free transitions, and data with much more limited spectral
resolution and S/N, Nicastro et al (1999) derived a density of
$7.4\times10^7$~cm$^{-3}$, and thus a distance $\sim$3 l-d for NGC4051.

Our $N_H$ values are factors of $\sim$20 (HIP) and $\sim$100 (LIP)
smaller than the one found by Nicastro et al (1999), who measured $log
N_H>$22.5. Our $n_e$ value for the HIP is also smaller by a factor 3-10. We
attribute  these differences to
three factors, which combined independently to cause these authors to
overestimate both quantities:

(1) They considered only 1 absorber to 
model the low-resolution ROSAT-PSPC data of NGC~4051.

(2) They estimated the absorber equivalent H column density by using a
spectral model that included only bound-free transitions, neglecting
the dominant contributions of resonant bound-bound absorption,
particularly by Fe ions (see their Fig. 4, also Krongold et al. 2003).

(3) Consequently, they used an empirical model to measure the
optical depths of the OVII and OVIII K absorption edges and find their
relative abundances (i.e. the ionization parameter of the gas). The
edges, however, coincide in energy with many Fe
bound-bound transitions. In contrast, we fit the higher quality 
XMM-{\em Newton} data of NGC~4051 with two absorbing components, and
use a spectral model (PHASE, Krongold et al 2003) that 
includes more than 3000 bound-free and bound-bound X-ray transitions from 
many metals (obtained mainly from the ATOMDB database, Smith et al. 2001)  to derive their equivalent H column 
densities. As a consequence, our determination of 
the relative abundances of OVII and OVIII is now much more
accurate than that of  Nicastro et al (1999). All $N_H$ values
determined from similar early WA models (Reynolds 1997, George
et al. 1998) will likewise be strong overestimates (see  Krongold et al 2003).

\section{The Accretion-Disk Origin of the Ionized Wind \label{loc}}
The sub-parsec-scale location of both absorbing components inferred
independently from high resolution RGS and low resolution EPIC data
can be used to test several models for the warm absorber: 

(1) X-ray observations of Seyfert~2 galaxies have shown the presence
    of extended, pc-kpc scale, bi-cones, of ionized gas responsible
    for the narrow X-ray emission lines detected in the X-ray spectra
    of these objects (Ogle et al. 2000; Sako et al. 2000; Sambruna et
    al. 2001;  Kinhkabwala et al. 2002; Brinkman et al. 2002; Ogle et
    al. 2003), and co-located, with the Narrow Emission Line
    Region. It has been suggested that the origin of the warm absorber
    is then at pc from the central source. Clearly, our results do
    not favor this possibility, as the origin of the wind in NGC~4051 is much
    further in. The narrow emission lines in Seyfert 2s should be associated
    with the narrow emission lines also observed in Seyfert 1s
    (e.g. Turner et al. 2003; Pounds et al. 2004), that do not vary in
    response to continuum changes in timescales of years, and thus
    are located much farther out than the absorber, at distances
    of pc, from the central source (Pounds et al. 2004 for NGC
    4051). Even though our results do not support the idea that the outflow starts
    at pc from the ionizing source, they are consistent with the possibility
    that the emission lines in both Seyfert 1s and Seyfert 2s are
    produced by the continuation of the same
    wind that forms the WA, as discussed in \S \ref{geomb}.

(2) It has been suggested that the structure of the warm absorbers in our   
line of sight is that of a continuous range of ionization structures
spanning a large radial region (of the order of pc) in the nuclear
environment, and spanning more than two orders of magnitude in
ionization parameter (e.g. Ogle et al. 2004, Steenbrugge et
al. 2005). However, in this model, no 
variability is expected in the opacity of the absorber with moderate flux 
variations (Krongold et al. 2005b). 
Both the changes in ionization state of the gas following the 
changes of flux observed in NGC 4051, and their sub-l-d 
thickness, rule out a continuous radial range of ionization stages, and
further support the idea of two distinct phases of a single medium for
the structure of the absorber.

\begin{figure}
\centering
\plotone{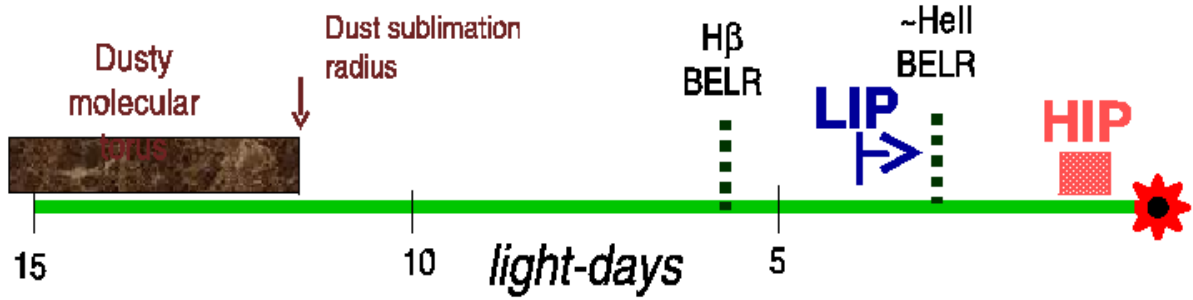}
\caption{Location of features in the nuclear environment
    of NGC 4051 on a light-day scale. \label{ld}}
\end{figure}

(3) The remaining non-disk origin for the ionized gas
-- evaporation off the inner edge of the dusty obscuring torus in AGNs
(Krolik \& Kriss 2001) -- is also ruled out
for both the HIP and the LIP (and again with both high and low
resolution data) in NGC 4051. The torus inner edge has to be
at a distance larger or equal to the dust sublimation radius: 
$r_{sub}=1.3L^{1/2}_{uv,46}T^{-2.8}_{1500}$ pc. Here $L_{uv,46}$ is
the UV luminosity of the source in units of $10^{46}$ erg s$^{-1}$
cm$^{-2}$ and $T_{1500}$ is the grain evaporation temperature in units
of 1500 K (Barvainis 1987). 
For NGC~4051 log$L(1450A)= 26.38 \pm 0.05$~erg~s$^{-1}$
Hz$^{-1}$ (Constantin \& Shields 2003) and, assuming a dust temperature
of 1500 K, we obtain $r_{sub}(NGC 4051)$= 3$\times$10$^{16}$cm
(0.01~pc, 12 l-d). 
This distance is a factor of 3.4 and 12 larger, respectively, than the LIP
and HIP upper limits. A hotter assumed dust temperature does not
remove the discrepancy
between the HIP and LIP location and the inner edge of the obscuring
torus as even for dust at a maximal temperature of 
2000~K, $r_{sub}$ is only reduced by a factor 0.45. Moreover, this
brings the sublimation radius to the same distance as the \hb\
BELR in NGC~4051, even
though the torus must be located outside the BELR for unification
models to work. Changing the UV luminosity also does not help if the X-ray
luminosity changes in parallel, as $r_{sub}$ and $R_{HIP}$ both scale
as $L^{0.5}$. Finally, this estimate of $r_{sub}$ is only a
lower limit for a putative torus-wind in NGC~4051. As Krolik \& Kriss
(2001) pointed out, the
innermost edge 
of a molecular torus, at which an ionized wind producing the absorption 
seen in the X-rays can form, is not simply at the sublimation radius. 
This edge must be derived using photoionization arguments for the 
observed gas, given the density of a putative torus wind (which is at
least 2 order of magnitudes lower than the densities we estimate here 
for the LIP and the  HIP). Following these arguments, Blustin et
al. (2005) derive a distance 
for the inner edge of the torus, of $r_{torus} \simeq 0.15$ pc for NGC
4051. This 
is 52 and 178 times larger than our upper limits on the LIP and HIP 
distances from the central ionizing source, respectively. Figure
\ref{ld} gives a schematic linear scale diagram of the distance in
light-days of the different AGN components to the central source.

Having ruled out all large-scale locations for the ionized wind, we
can look more carefully at the possible small-scale locations.
The size of the H$\beta$ BELR in NGC~4051, from reverberation mapping is
5.9~l-d (Peterson et al. 2000). As
$R_{HIP}=$0.5-1.0~lt-days, the HIP is 
clearly much smaller than the H$\beta$ BELR. The LIP location, on the
other hand, is marginally 
consistent with the H$\beta$ BELR. The  higher ionization HeII broad
emission line has a smaller measured reverberation lag $\simlt$1 l-d
(Peterson et al. 2000), consistent with the HIP
location. This broad line  has a
wing blueshifted by $\sim$400~km~s$^{-1}$ compared to \hb, suggesting a 
wind component, which in turn suggests a connection with the X-ray warm 
absorber  (which has outflow velocities of 600-2340~km~s$^{-1}$, e.g.
Collinge et al. 2001). Peterson et al. (2000) measure
FWHM(HeII)= 5430~km~s$^{-1}$
(in the rms spectrum). If the HeII gas were virialized, the HeII
BELR is $\sim$2 times closer in than the \hb\ BELR, i.e. $\sim$2.5~l-d, which is
consistent with the largest allowed LIP size. The HeII outflow, however,
implies a somewhat narrower virial component to the line and so a larger
distance from the black hole. An accurate measurement of the HeII
reverberation lag would give a definitive answer.

\begin{figure}
\centering
\plotone{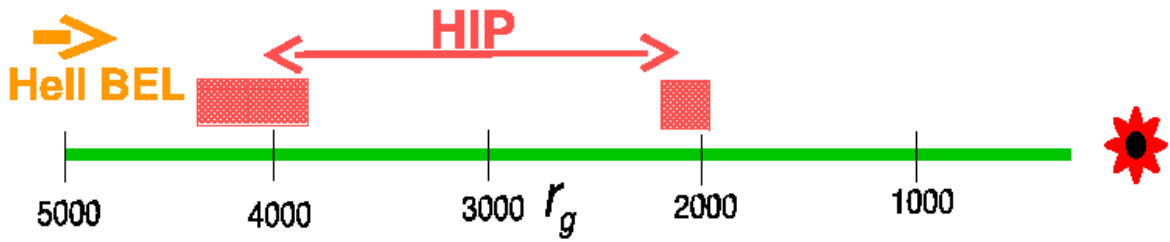}
\caption{Location of features in the nuclear environment
    of NGC 4051 on a gravitational radius scale. \label{sr}}
\end{figure}

The location of the HIP in Schwarzschild radii ($R_s$) implies that
the origin of the HIP is related to the accretion disk.
In these units, $R_{HIP}$= 2300 - 4600~$R_s$, which locates the 
outflow in the outer regions of the accretion disk, but at distances
shorter than the radius where an $\alpha$-disk 
becomes gravitationally unstable ($\sim35000~$R$_s$; Goodman 2003). The LIP is also 
likely related to the disk, as $R_{LIP}<15800R_s$.  Figure
\ref{sr} gives a scale diagram of the location of the different AGN components.
We note that NGC~4051 has a fairly low accretion rate relative to
Eddington value (10\%, Peterson et al. 2004) for a Narrow Line Seyfert 1
Galaxy (Pounds et al. 1995). This could be because we are
seeing it in a pole-on configuration and so this gives narrower
broad emission lines than normal. If so then the black hole mass for
this object is underestimated, and the location of the absorbing
components in $R_s$ should decrease, making a more compact absorber.

\section{Geometry of the Wind \label{geomb}}

Our findings considered alone, are consistent with thin spherical
shells of material 
which are expelled radially from the central ionizing source. However, 
we consider this configuration implausible because of the fine tuning 
required in the frequency with which these shells of material would
have to be produced to explain the high occurrence of
WAs in AGNs, and yet still avoid having the thin shells degenerate into a
continuous flow (which is strictly ruled out for NGC~4051). 
The next simplest geometrical configuration is that of a 
bi-conical wind (Fig. \ref{gwind}, Elvis 2000). All our 
estimates for the physical and geometrical quantities of the two X-ray 
absorbers of NGC~4051 are consistent with this scenario, and recent
magneto-hydrodynamical simulations also favor this interpretation
(Everett 2005).

We note that assuming this bi-conical geometry, our results are
consistent with the possibility that the extended, pc-kpc scale,
emitting bi-cones
observed in  Seyfert 2s are part of the same flow that forms the WA
(e.g. NGC 1068, Crenshaw \& Kraemer 2000).
In this case the NELR and X-ray emitting gas would be the continuation of the
WA wind but at large radii. A constant ionization parameter with
radius could  be expected if the density decreases with the
inverse square of the distance, as observed by Bianchi et al. (2006)
for the kpc scale nebulae in type 2 objects.

The possible connection between these two components requires that the
emitting bi-cones are not filled with material (as otherwise, from a
pole-on line of sight we should observe an absorber forming a
continuous radial range of ionization stages, which is already ruled out). Then
from a pole-on configuration no absorption would be observed, as no
material would cross our line of sight  (Fig. \ref{gwind})  in
agreement with the detection rate of WA.  
The absorber would be observed as a transverse flow (which is required
by observations, see below), through the inner and outer edges of the
bi-cone, and
close to the base of the flow, as shown in Figure \ref{gwind}.
The observed X-ray emission would be produced by the same bi-conic
flow, but at much larger distances. In
this scenario absorption and emission are then part of the same wind,
although the two measured distances are very different. This idea agrees
with recent results on Seyfert 2 galaxy NGC 1068 by Das et al. (2006),
who found that the best kinematic model for the X-ray bi-conical
emission requires empty cones, with a geometry consistent to the one
found here for the warm absorber.

\section{Wind Mass and Kinetic Energy Outflow Rate \label{sec:mass-out}}

With the location and density of the X-ray absorbing material in hand,
and assuming the above bi-conical geometry, we can now estimate the mass outflow
rate of the wind.

The escape velocity ($v_{esc}$) from the location of the HIP is $\sim$4000-6000
km s$^{-1}$, which is 10 times larger than the measured outflow
velocities in our line of sight for the X-ray warm absorber in NGC 4051
(v$_{r}\sim$ 500 km s$^{-1}$). The true outflow velocity is most
likely to be 2 times larger (i.e. for a 60 degree angle between our
line of sight and the wind). This, however is still $\sim$5 times
smaller than the escape velocity at the distance of the HIP. Either 
the wind falls back, or the wind is accelerated after it crosses our
line of sight. If the gas were falling back,
evidence of inflowing material in absorption should be detected
(in the X-ray band or in the UV band through the Ly$-\alpha$ line if
the gas has cooled down), and redshifted absorbing systems are
rare and do not necessarily imply inflow (e.g. Vestergaard 2003).
Since there is evidence of a transverse wind across our
line of sight in other AGNs (Mathur et al. 1995; Crenshaw, Kraemer \&
George 2003) which is accelerating (Arav 2004), the latter option is
preferable. In addition, radiatively line driven wind
models (Proga \& Kallman 2004) and phenomenological derived models
(Elvis 2000), require later acceleration of the wind. Furthermore,
in a conical wind, the material of the warm absorber must be rotating with
Keplerian velocity ($v_{Kep}$)
around the center (at least at the base of the wind). Since the escape
velocity scales with $v_{Kep}$ as $v_{esc}=\sqrt{2}~v_{Kep}$, the wind
already has half the kinetic energy needed to escape. All these point
to an escape of the material from the central regions. 

In a bi-conical wind, from a direction looking directly along the cone
opening angle an
observer would see a much larger column density, $>1.5\times10^{23}$
cm$^{-2}$ (assuming the wind can reach $v_{esc}$) mostly due to the
LIP (see Appendix A.1). This high N$_H$ flow would
also be seen to have a larger velocity, perhaps much larger, as in WA we may be
seeing the flows before they get fully accelerated.  Thus, from this
special direction the NGC~4051 wind would begin to resemble a `Broad Absorption Line' (BAL) system (Elvis 2000).

To evaluate the mass loss rate in the WA wind let $\delta$ be the
angle between our line of sight to the central 
source and the accretion disk plane, and $\phi$ be the angle formed by the wind with 
the accretion disk (see Fig. \ref{gwind}). We can then derive a
formula for the mass outflow 
rate and write it in terms of the observables $v_r$ (the line of 
sight outflow velocity), N$_H$ (the line of sight equivalent H column density), and
$R$ (see Appendix A.2 for full details on the derivation of this
formula):
\begin{equation*}
\dot{M}_w = 0.8 \pi m_p N_H v_r R f(\delta,\phi)
\end{equation*}
where $f(\delta,\phi)$ is a factor that depends on the particular
orientation of the disk and the wind, and for all reasonable angles
($\delta> 20^o$ and $\phi>45^o$)
is of the order of unity (with a variation by a factor of 2, see Fig. \ref{angles}). 
Thus, for a vertical disk wind ($\phi = \pi/2$) and an average Seyfert 1-like 
line of sight angle $\delta = 30^0$  $f(\delta,\phi)$=1.5 (see A.2), and we
find: $\dot{M}_w(HIP)=(0.7-1.4) \times 10^{-4}$ M$_{\odot}$ yr$^{-1}$
and  $\dot{M}_w(LIP) < 0.9 \times 10^{-4}$ M$_{\odot}$ yr$^{-1}$. 
For NGC~4051, assuming 10\% accretion efficiency, the observed accretion 
rate is $\dot{M}_{accr} = 4.7 \times 10^{-3}$ M$_{\odot}$ yr$^{-1}$
(Peterson et al. 2004), 
so $\dot{M}_w(HIP) = (0.02-0.03) \dot{M}_{accr}$ and $\dot{M}_w(LIP) < 0.02$ 
$\dot{M}_{accr}$. 
Therefore, the total mass outflow rate from the LIP and the HIP, in NGC 4051, 
is $ \dot{M}_{out} = (0.02-0.05) \dot{M}_{accr}$. Thus, the total mass
outflow rate is only a small fraction 
of the observed accretion rate. 
We stress here that these estimates of the mass outflow rate from the X-ray 
disk winds of NGC~4051 depend only weakly on the assumed 
geometrical configuration (unless NGC~4051 is seen close to pole-on,
see end of \S \ref{loc}). Their robustness is due to the 
strong constraints that we can independently set on
the physical properties of the absorbers.


\section{Possible Implications for Galaxy Evolution and Cosmic Feedback}

Given the importance that recent theoretical studies have bestowed
on AGN winds (see below), it is worth investigating the potential implications that warm
absorber winds can have on their large scale environment. Some aspects
of this discussion are speculative and we do not pretend to present a physical
model. Rather, we present only an analysis in terms of energy budgets
and order of magnitude estimates.

Assuming that the black hole mass of NGC 4051 was all accreted
(M$_{BH} = 1.9 \times 10^6$ M$_{\odot}$,  Peterson et al. 2004), we
can calculate the integrated lifetime mass lost due to the X-ray winds,
M$_{out} = (0.4-1.0) \times 10^5$ $M_{\odot}$ (M$_{out}= 0.02-0.05\times$
M$_{BH}$). This number is too small for the winds to have an important
influence on the ISM or IGM. However, NGC 4051 is on
the low end of AGN black hole masses,
luminosities ($L_{bol} = 2.5 \times 10^{43}$ \ergss, Ogle et al. 2004) and accretion
rates relative to Eddington value (10\%, Peterson et
al. 2004). If the measured ratio 
$\dot{M}_{out}/\dot{M}_{accr}$ of NGC 4051 is representative of the quasar
population, then for winds in powerful quasars ($L_{bol} \sim 10^{47}$
\ergss, M$_{BH} = $ few$\times10^9$ M$_{\odot}$) the total mass
outflow could be as large as
$M_{out} \sim$~few$\times10^8$ M$_{\odot}$ (neglecting any additional
mass from the interstellar medium entrained in the wind). This mass is
comparable to that available from the most optimistic estimates of
powerful ULIRGs' winds (which include significant mass entrainment,
e.g. Veilleux et al. 2005) and could, in principle, be deployed into 
the IGM surrounding the quasar's host galaxy.

The quasar nuclear environment is unusually rich in metals, with
metallicities several
times solar found in winds from local AGN. Fields et al. (2005) found
metallicities 5 times solar in Mrk 1044, and higher values are
likely for high z, high luminosity quasars (e.g. Hamman \& Ferland 1999). 
On the other hand, metallicities in Lyman $\alpha$ forest at $z\sim 3$ are
low (0.01-0.001 times solar, Pettini 2004). So, assuming
that quasar winds can escape the host galaxy, they
could feed their local
IGM with highly metal enriched material. This would create local
inhomogeneities in the metal content of the IGM around such quasars.    
From our line of sight to powerful quasars, the measured column
density of metals in IGM systems close to the central object (at
distances up to a few hundred kpc) would then be
much larger ($>$100) than in Lyman $\alpha$ systems located far away
from quasars.
Thus, metal rich IGM systems may be significantly stronger close to
powerful quasars.

Simulations show that, if quasar winds are to
be the main process controlling the evolution of the host galaxy and
the surrounding IGM, they require
output kinetic energies of the order of  $\sim10^{60}$ erg
(e.g. Hopkins et al. 2005; Scannapieco 2004; King
2003). Such energy outputs would be enough to produce the well known
relation between the central black hole mass and bulge velocity
dispersion (Ferrarese \& Merrit 2000; Gebhardt et al. 2000), or to
heat the IGM controlling structure formation. 

However, lower wind energy outputs may still have an important effect on
their environments. For instance, unbinding the hot phase of the ISM
may be enough to stop large scale star formation processes in a host
galaxy. As this
confining medium escapes from the galaxy, the warm and cold phases
must expand, decreasing their temperature and density, and producing a fast decline
in the star formation rate (a full analysis will be presented in a
forthcoming paper, see also Natarajan et al. 2006). Evaporating the
hot-ISM from a typical galaxy requires that
the wind heats this medium from its current
temperature of $\sim10^6$~K to $\sim10^7$~K. The energy needed
to increase the gas temperature by this amount is E$\sim$N$_{TOT}k$T
(where N$_{TOT}$ is the total number of particles in the disk and
T$\sim10^7$ K). Assuming a galactic disk of
radius 10 kpc and thickness 0.1 kpc, with a hot-ISM electron
density of $10^{-2}$~cm$^{-3}$ (Smith et al. 2006), implies
that an energy input of $\sim10^{55}$ erg into the hot-ISM would heat it
to T$\sim10^7$ K and would 
produce this medium to evaporate, ceasing star formation.   
For the observed velocities found in WAs ($\sim$ 500 km
s$^{-1}$), in $10^7$ yr the gas would move 5 kpc away from
the central region, into the host galaxy. This time is consistent with
the lifetime inferred for quasars ($10^7$-$10^8$ yr), and suggests
that the wind could indeed affect the host galaxy. 
The total kinetic power released by
the wind in NGC 4051 (given by
$\dot{M}_{out}v^2/2$) is $\sim 10^{38}$ erg s$^{-1}$ (assuming the
observed outflow velocity of 500 km s$^{-1}$), and the total
kinetic energy deployed into the ISM/IGM surrounding NGC 4051 would be only
$\sim10^{54}$ erg. This energy is still an order of magnitude smaller
than the $10^{55}$ erg required to disrupt the hot phase of the
ISM. However, even WA winds in typical Seyfert 1 galaxies (with black hole
masses few$\times 10^7$ M$_{\odot}$) would reach this energy, and
could have an effect on the hot-ISM and star formation processes of
their hosts.

If the measured ratio
$\dot{M}_{out}/\dot{M}_{accr}$ is representative
of quasars, then (assuming velocities in our line of sight of
500 km s$^{-1}$) the total energy deployed by a powerful quasar into
the host galaxy ISM or the IGM, during the AGN active
phase, could be $\sim$ $few \times10^{57}$ erg. Thus, winds in powerful
quasars could have important effects controling star formation in
their hosts (disrupting the hot-ISM), but  would still be
too weak to have a drastic influence on the whole galaxy evolution.
We suggested, however, that
we are not seeing the full terminal velocity of
the wind (\S \ref{sec:mass-out}; Proga \&
Kallman 2004; Elvis 2000). Then, if
the winds are accelerated to v$\sim$5000 km s$^{-1}$, i.e. by a factor of
10 (which is the ratio between the velocities found in WAs and BALQSOs),
the total kinetic energy deployed by these systems could reach   
$\sim$ few$\times10^{59}$ erg, similar to the binding energy of
massive galactic bulges.
This energy is comparable to the 10$^{60}$ erg
required by simulations for quasar winds to be critically
important in feedback processes (e.g. Hopkins et al. 2005; Scannapieco
2004; King 2003).





Thus, despite our low mass loss rate estimates, it still remains
possible that AGN winds are important for at least two major feedback
mechanisms: (1) the evolution of their host galaxies, putting energy
into the ISM of the host and controlling the accretion 
process, thus regulating the black hole growth (the 
relationship between the mass of the central black hole in galaxies
and the velocity dispersion of the galactic bulge),
and/or (2) enriching with metals the IGM and heating this medium,
controlling the accretion of
material onto galaxies (thus controlling structure
formation). Furthermore, the above conclusions are based on a one-object
analysis.  The ratio
$\dot{M}_{out}/\dot{M}_{accr}$ could well be larger for quasars,
increasing the mass and energy output of quasar winds.  To fully
determine what role do AGN winds play require measuring AGN geometry
and kinematics for more normal objects, preferably over a range of
black hole masses and Eddington ratios.

\section{Conclusions}
The response of the ionization state of the
absorbing gas in NGC~4051 to variations in the ionizing continuum
immediately rules out a radially continuous flow and a large scale (kpc)
origin for the WA (at the location of the NLR). With a detailed analysis we
were able to constrain the density and
location of the absorbing gas. 
The two-component absorber in NGC 4051 is dense
[$n_e =(5.8-21.0) \times 10^6$~cm
$^{-3}$ for the high ionization absorber and $n_e >
$8.1$\times$10$^7$~cm$^{-3}$ for the low ionization one], and its location
is close to the continuum source, at 
distances 0.5-1.0~l-d (2200 - 4400R$_s$) and $<$ 3.5~l-d
($<~15800$R$_s$) for the high and low ionization components, respectively.
The two absorbing components are in
pressure balance strongly suggesting a two-phase medium.
These results for NGC~4051 place this WA wind on an accretion disk
scale, well within the inner edge of a dusty torus or the low
ionization H$\beta$ broad emission line region. 
A continuous radial wind with varying ionization state is doubly
ruled out, as the absorbers have a thickness $\Delta$R $<<$ R. This
thinness also makes purely radial
motion unlikely, given the 50\% incidence of ionized absorbers in AGNs.
A bi-conical outflow largely transverse to our line-of-sight is the
simplest geometry that can produce the small relative thickness of the
absorbers and their high covering factor.
Since the smallest radius
found is the one that gives the strongest constraint on the
origination radius of WAs, we conclude that AGN winds must arise from the accretion disk and
points us toward physical models for the production of winds from AGN disks.

An accretion disk wind can tie together all the absorption and
emission features in the spectra of AGN, including the broad emission
lines (Elvis 2000). The location of the He II BELR is consistent with
the position of the absorber, supporting the above
picture. Our results are also consistent with the idea that the
pc, bi-conical, emission regions observed in Seyfert 2 galaxies are the
continuation, at much larger distances from
the central region, of the warm absorber winds.

The implied mass outflow rate of the warm absorber wind in NGC 4051 is
a small fraction of the mass accretion rate ($ \dot{M}_{out}$ = 2-5\%
$\dot{M}_{accr}$). If all the mass in the central black hole of NGC
4051 was accreted, then the total outflow of mass produced by the wind
would be M$_{out} = (0.4-1.0) \times 10^5$ $M_{\odot}$ and the
total kinetic energy released by the flow would be $\sim10^{54}$
erg. This energy is too small to unbind the entire ISM of the host galaxy in
this AGN, but is comparable to the energy required to disrupt the hot phase
of the ISM, which could then disrupt star formation.
This suggests that even mild winds in Seyfert galaxies might have an important
effect in their host galaxies. 

Only through fully understanding AGN outflows, we will be able to
calculate AGN feedback accurately.
Further studies of less extreme AGNs over a large range of black hole
masses and accretion rates relative to Eddington will give a
definitive answer.

\acknowledgements 
We thank the referee for constructive comments that helped to improve
the paper. This research is based on observations obtained with XMM-Newton,
  an ESA science mission with instruments and contributions directly
  funded by ESA Member States and NASA. This work was supported by the
  UNAM PAPIIT grant IN118905 and the CONACyT grant 40096-F.
 N. Brickhouse was supported by NASA contract NAS8-39073 to the
 Chandra X-ray Center. F. Nicastro acknowledges support from NASA LTSA
 grant NNG04GD49G.


\appendix
\section{Appendix}

\subsection{The Column Density of a Disk-Wind observed directly down the cone}

In this section we derive the column density of a conical wind viewed
down the cone. We assume that the cone bends forming a funnel-shaped
geometry. In such geometry, after the point when the flow bends, it
can be considered again as a radial flow. From the equation of mass
conservation it follows that
\begin{equation}
n_H(R) = n_{oH}\times (v(R)/v_o)^{-1} \times
(R/R_o)^{-2}
\eqnum{1}
\end{equation}
where the subscript $o$ stands for the
conditions at the base of the radial flow, i.e. the bending point of
the flow in the funnel-shape geometry. We assume that the
properties at the base of the radial flow are equal to the measured properties
in our analysis. 

Now consider two forces acting on the gas: the gravitational attraction
of the central black hole, and the radiative force on the warm
absorber. Since both forces are proportional to $R^{-2}$, we can
simply write the equation of motion for the warm absorber as:
\begin{equation}
M_{out}\frac{dv(R)}{dt} = \frac{k}{R^{2}}
\eqnum{2}
\end{equation}
with $k$ constant. It can be easily shown that for an object with luminosity
L$\sim$0.05 L$_E$ radiative force will dominate over gravitacional
force. Thus for most Seyfert 1 galaxies and quasars $k>$0.
Solving equation (2) for $V(R)$ gives
\begin{equation}
\frac{v(R)}{v_o} = \left[ 1 + \frac{2k}{v_o^2R_oM_{out}}
\left(1-\left[\frac{R}{R_o}\right]^{-1}\right) \right]^{\frac{1}{2}}
\eqnum{3}
\end{equation}
From the equation of energy conservation for the flow
\begin{equation}
\frac{1}{2}M_{out}v_{\infty}^2=\frac{1}{2}M_{out}v_{o}^2+\frac{k}{R_o}
\eqnum{4}
\end{equation}
where the subscript $\infty$ stands for the final velocity of the
wind. It can be immediately derived that 
\begin{equation}
\frac{2k}{v_o^2R_oM_{out}} = \left(\frac{v_{\infty}}{v_{o}}\right)^2
- 1 
\eqnum{5}
\end{equation}
Replacing equation (5) in (3) gives
\begin{equation}
\frac{v(R)}{v_o} = \left[ 1 + \left(\left(\frac{v_{\infty}}{v_{o}}\right)^2
- 1\right) \left(1-\left[\frac{R}{R_o}\right]^{-1}\right) \right]^{\frac{1}{2}}
\eqnum{6}
\end{equation}

Replacing equation (6) into equation (1) and integrating for the column density
$N_H=\int n_(r)dr$ (where the integral runs from $R_o$ to $\infty$)
it is found that
\begin{equation}
N_{\infty}=\frac{n_{o}R_o}{1+(v_{\infty}/v_o)}
\eqnum{7}
\end{equation}
Thus the column density only
depends on the values at the base of the radial flow. 

We assume that the conditions measured for the LIP and the HIP in NGC 4051 are
representative of the conditions at the base of the radial flow. Thus we
assume that $n_{o}(H)$ is equal to the value obtained in our analysis,
that $v_o$ is equal to the observed line of sight velocity of the flow
and that $v_{\infty}=v_{esc}$. With this conditions, and assuming
$(v_{esc}/v_o)\sim10$ (if the flow viewed directly down
is observed as a BALQSO) then
$N_H(LIP) >1.5\times10^{23}$ cm$^{-2}$ and $N_H(HIP)\sim
4\times10^{21}$ cm$^{-2}$. So, the column density in this case is
dominated by the LIP, and the total equivalent H column density measured
down the flow should be $N_H(LIP) >1.5\times10^{23}$  cm$^{-2}$.

\subsection{The Mass Outflow Rate of a Disk-Wind in a conical Geometry}

\begin{figure}
\plotone{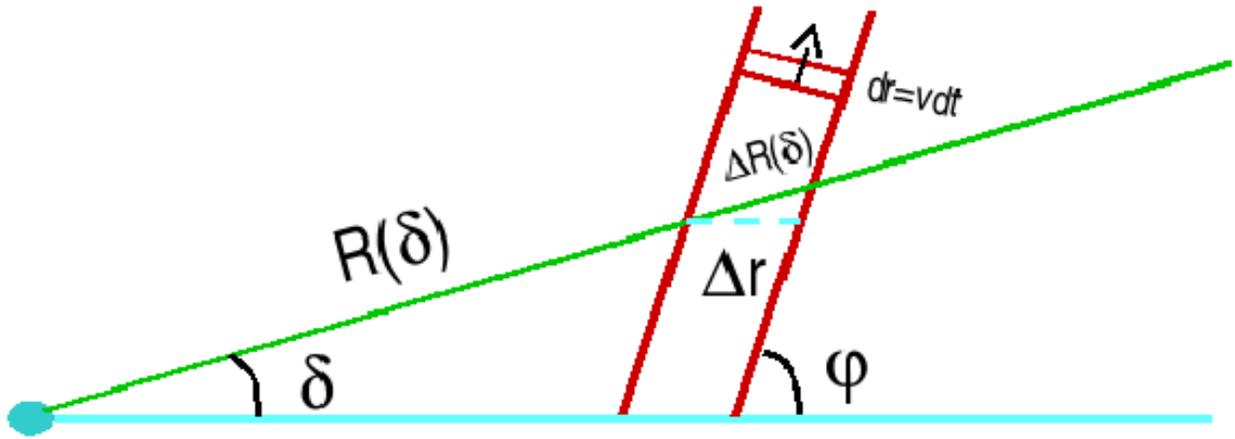}
\caption{Disk-Wind in a conical geometry. 
$\phi$ is the angle formed by the wind and the disk and $\delta$ the
  angle between the disk and the observer's line of sight. $R$ is the
  distance from the continuum source to the wind and $\Delta~R$ its
  thickness. \label{gwind}}
\end{figure}

\begin{figure}
\plotone{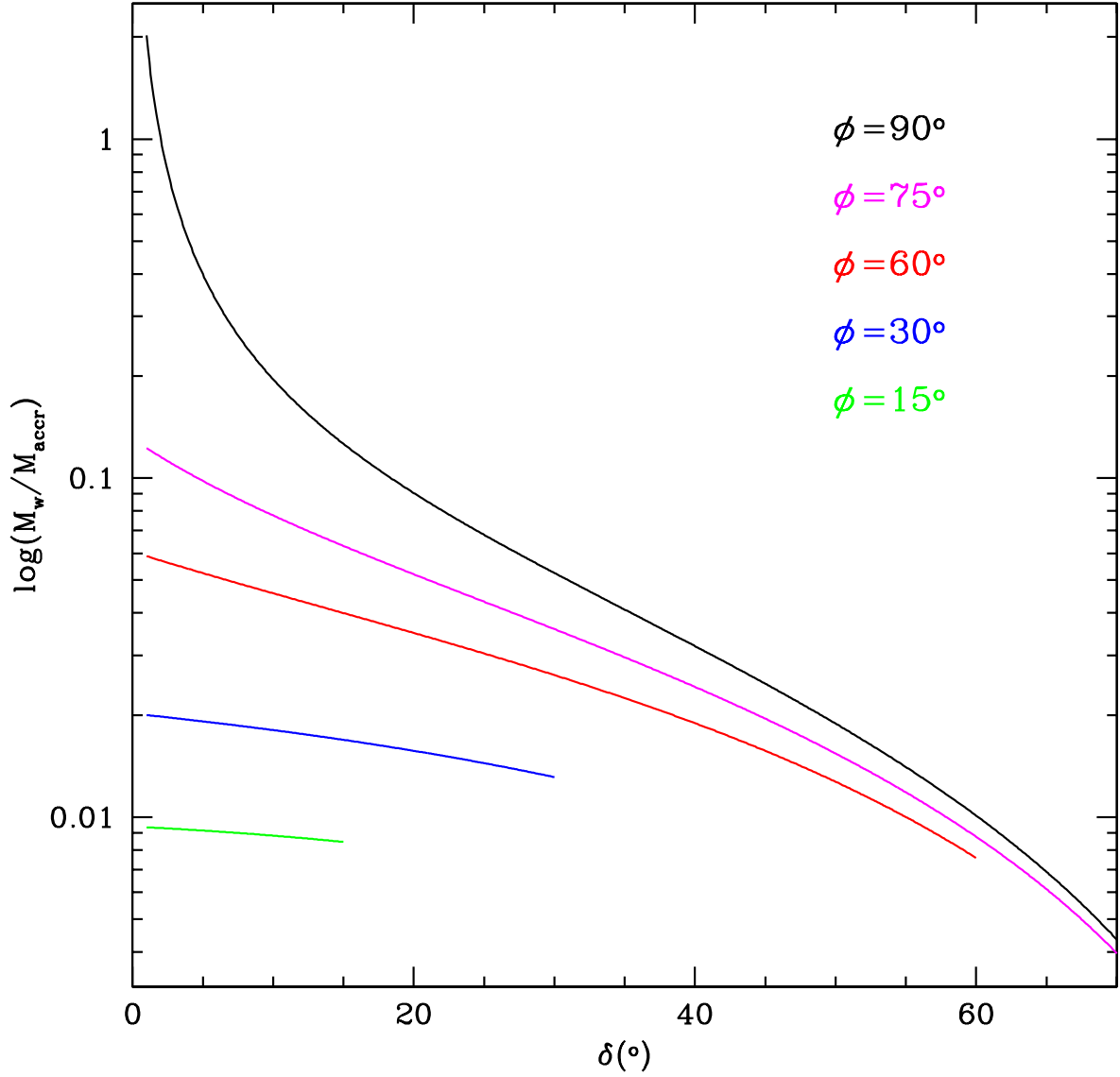}
\caption{Mass outflow rate (in accreted mass units) vs. line of sight angle
($\delta$) for several values of the inclination of the wind 
($\phi$). For reasonable angles the mass outflow only changes by a
factor $\sim 2$.\label{angles}}
\end{figure}

Let $\delta$ be the angle between our line of sight to the central 
source and the accretion disk, and $\phi$ the angle formed by the wind with 
the accretion disk, as shown in Figure \ref{gwind}. 
If $r$ and $\Delta r$ are the projections of the line of sight distance of the 
wind from the central source ($R(\delta)$) and the line of sight 
{\em effective}\footnote{Here by {\em effective} we mean the net observable thickness of 
the gas, allowing for clumping in the flow. Thus, in our treatment we do  
not include a linear (or volume) filling factor, since we are interested 
in estimating the net flow of mass, starting from the observables.}
thickness of the wind ($\Delta R(\delta) = 1.23 N_H / n_e$) onto 
the accretion disk plane, then $r = R cos\delta$ and $\Delta r = \Delta R 
cos\delta$. 
Let us further suppose that the true thickness $\Delta R = \Delta r sin 
\phi$ of the wind (between the two conical surfaces) stays constant with 
$\delta$ (i.e. $\Delta r \neq \Delta r(\delta)$). 
With these relations we can now estimate the net mass loss rate in the 
outflow. 
In general, $\dot{M}_w = (1/1.23) n_e m_p \times v_r / sin(\phi - 
\delta) \times A$, where, $v_r$ is the line of sight component of the 
outflow velocity and $A$ is the area (constant with $\delta$) effectively 
filled by the gas and defined by the projection of the section of two 
concentric cylinders perpendicular to the accretion disk plane, with inner 
and outer radii $r$ and $r + \Delta r$, onto the plane perpendicular to the 
outflow velocity: $A = \pi [(r + \Delta r)^2 - r^2] sin
\phi$. Rearranging this expression 
in terms of observables, $\dot{M}_w = 0.8 n_e m_p \times v_r / sin(\phi - 
\delta) \times \pi [(R + \Delta R)^2 - R^2] cos^2\delta sin \phi = 
0.8 (n_e R^2) m_p [v_r / sin(\phi - \delta)] \times \pi [(\Delta R / R)^2 + 
2 (\Delta R / R)] cos^2\delta sin \phi$. This formula can be
simplified to $\dot{M}_w = 0.8 \pi m_p N_H v_r R f(\delta,\phi)$ (where
$f(\delta,\phi)$ collects the angle dependences), since
$(\Delta R / R)^2$ is negligible due to the small relative thickness
of the absorbers.

Thus, for a vertical disk wind ($\phi = \pi/2$) and an average Seyfert 1-like 
line of sight angle $\delta = 30^0$ we find: $\dot{M}_w(HIP)=(0.7-1.4)
\times 10^{-4}$ M$_{\odot}$ yr$^{-1}$ and  $\dot{M}_w(LIP) < 0.9 \times 
10^{-4}$ M$_{\odot}$ yr$^{-1}$. 
For NGC~4051 $L_{BOL} \sim 2.7 \times 10^{43}$ erg s$^{-1}$ (Ogle et
al. 2004). Assuming 10\% accretion efficiency
the observed accretion rate for NGC~4051 is $\dot{M}_{accr} = 4.7 
\times 10^{-3}$ M$_{\odot}$ yr$^{-1}$. 
Thus the mass outflow rates, in terms of mass accretion, are
$\dot{M}_w(HIP) = (0.02- 0.03) \dot{M}_{accr}$ and $\dot{M}_w(LIP) < 0.02$ 
$\dot{M}_{accr}$. This outflow rates cannot change much (factors $\sim
2$) for different (reasonable) choices of $\delta$ and $\phi$
($\delta> 20^o$ and $\phi>45^o$), as can
be seen from Figure 16.
%


\clearpage
\begin{deluxetable}{lccc}
\tablecolumns{4} \tablewidth{0pc} \tablecaption{Narrow Emission Lines used in the models. \label{tabem}}
\startdata
\hline
Line       & $\lambda_{source}$ & Width (km s$^{-1}$) & Flux
  (10$^{-5}$ erg cm$^{-2}$ s$^{-1}$) \\ 
\hline 
C VI Ly$_\alpha$  & 33.79$\pm$0.03 & 358$\pm$89     &  5.5$\pm$1.5 \\ 
N VI 1s-2p (f)    & 29.59$\pm$0.03 & 137$\pm$31     &  2.5$\pm$1.0 \\
O VII 1s-2p (f)   & 22.14$\pm$0.03 & 194$\pm$37     & 12.0$\pm$2.5 \\
O VII 1s-2p (i)   & 21.84$\pm$0.03 & 194$\pm$37$^{a}$ &  5.5$\pm$1.5 \\
O VII 1s-2p (r)   & 21.65$\pm$0.06 & 194$\pm$37$^{a}$ &  3.0$\pm$1.5 \\
O VIII Ly$_\alpha$& 19.01$\pm$0.02 & 134$\pm$21     &  6.5$\pm$1.0 \\
Fe XVII 2p-3s     & 17.10$\pm$0.02 & 397$\pm$93     &  2.5$\pm$1.0 \\
Ne IX 1s-2p (f)   & 13.73$\pm$0.02 &  98$\pm$21     &  3.0$\pm$1.0 \\
\hline
\enddata

\tablenotetext{a}{Constrained to the best fit value of the O VII 1s-2p (f) line.}
\end{deluxetable}

\clearpage
\begin{deluxetable}{lcccccc}
\tablecolumns{7} \tablewidth{0pc} \tablecaption{Best-fit values over
  spectra extracted for the whole, time integrated,  XMM-Newton RGS and
  EPIC observation.
  \label{tab1}}
\startdata
\hline
RGS & $\Gamma$ & BB1(kT) (keV) & BB2(KT) (keV) & log(N$_H$) [cm$^{-2}$] &
log(U$_x$) & vel. (km s$^{-1}$) \\ 
\hline
HIP &   1.74$\pm0.5$ & 0.137$\pm{.06}$ &  0.064$\pm{.08}^a$   & 21.42$\pm{.12}$ &
-0.76$\pm{.08}$ & 537$\pm{130}$ \\
LIP &   1.74$\pm0.5$ & 0.137$\pm{.06}$ &   0.064$\pm{.08}^a$  &
20.73$\pm{.17}$ & -2.90$\pm{.10}$ & 492$\pm{97}$ \\
\hline
EPIC & $\Gamma$ & BB1(kT) (keV) & BB2(KT) (keV) & log(N$_H$) [cm$^{-2}$] & log(U$_x$) & vel. (km s$^{-1}$)\\ 
\hline
HIP &   1.97$\pm0.09$ & 0.141$\pm{.03}$ &   0.064$\pm{.08}$  & 21.42$\pm{.12}$$^b$ &-0.72$\pm{.08}$ & 537$\pm{130}$$^b$ \\
LIP &   1.97$\pm0.09$ & 0.141$\pm{.03}$ &   0.064$\pm{.08}$  &
20.73$\pm{.17}$$^b$ & -2.97$\pm{.14}$ & 492$\pm{97}$$^b$ \\
\hline
\enddata

\tablenotetext{a}{Constrained to the best-fit value on the EPIC model.}
\tablenotetext{b}{Constrained to the best-fit value on the RGS model.}
\end{deluxetable}

\clearpage

\begin{deluxetable}{lcccc}
\tablecolumns{5} \tablewidth{0pc} \tablecaption{RGS and EPIC-PN models for the HS and the LS. \label{tab3}}
\startdata
\hline
Abs. & RGS (HS) & RGS (LS) \\
\hline
HIP log(U$_x$) & $-0.72_{-.05}^{+.08}$ & $-1.31_{-.25}^{+0.38}$ \\ 
LIP log(U$_x$) & $-2.91_{-.11}^{+.10}$ & $-3.65_{-.50}^{+.25}$ \\
\hline
Abs. & EPIC (HS) & EPIC (LS) \\
\hline
HIP log(U$_x$) & $-0.68_{-.04}^{+.09}$ & $-1.24_{-.11}^{+.16}$ \\
LIP log(U$_x$) & $-2.98_{-.10}^{+.15}$ & $-3.72_{-.17}^{+.17}$ \\ 
\enddata
\end{deluxetable}

\clearpage

\begin{table}
\caption{Physical parameters of High and Low Ionization Absorbers \label{tab2}}
\footnotesize
\begin{tabular}{lcccccccc}
\hline
Abs. & N$_H$ & T$_e$ & $(n_eR^2)$ & $n_e$ & P$_e$ & $R$ & 
$\Delta R$ & $(\Delta R / R)$ \\
& $10^{21}$ cm$^{-2}$ & $10^5$ K & $10^{38}$ cm$^{-1}$ & $10^{7}$ cm$^{-3}$ &
$10^{12}$ K cm$^{-3}$ 
& $10^{15}$ cm & $10^{14}$ cm & \\
\hline
HIP & $(3.2 \pm{2.2})$ & $(5.4 \pm{1.4})$ & $(0.38 \pm 0.05)$ & (0.58-2.1) & (2.9-10.5) & 
(1.3-2.6) & (1.9-7.2) &  (0.1-0.2) \\
LIP & $(0.59 \pm{0.28})$ & $(0.28 \pm{.04})$ & $(66 \pm 3)$  & $>8.1$ & $> 2.4$ & 
$< 8.9$ & $< 0.09$ & $< 10^{-3}$ \\
\hline
\end{tabular}
\end{table}

\clearpage


\clearpage

\clearpage



\clearpage

\clearpage

\clearpage

\clearpage

\clearpage

\clearpage

\clearpage

\clearpage

\clearpage

\clearpage
\clearpage

\end{document}